\documentstyle[epsf]{elsart}  
\begin{document}
\newcommand{\beq}{\begin{equation}}
\newcommand{\eeq}{\end{equation}}
\newcommand{\beqa}{\begin{eqnarray}}
\newcommand{\eeqa}{\end{eqnarray}}
\newcommand{\Ss}{{\tilde S}}
\newcommand{\Vs}{{\tilde V}}
\newcommand{\Ts}{{\tilde T}}
\newcommand{\As}{{\tilde A}}
\newcommand{\Ps}{{\tilde P}}
\newcommand{\PVs}{{\widetilde {PV} }}
\newcommand{\fps}{f_{\pi}^2 }
\newcommand{\mks}{m_{{\mathrm K}}^2 }
\newcommand{\ms}{m_{{\mathrm K}}^{*} }
\newcommand{\msq}{m_{{\mathrm K}}^{*2} }
\newcommand{\rhos}{\rho_{\mathrm s} }
\newcommand{\rhob}{\rho_{\mathrm B} }
\newcommand{\kf}{k_{\mathrm F} }
\newcommand{\Sigs}{\Sigma_{\mathrm s} }
\newcommand{\Sigv}{\Sigma_{\mathrm v} }
\newcommand{\Sigo}{\Sigma_{\mathrm o} }
\newcommand{\mst}{ {\tilde M}^* }
\newcommand{\mstq}{ {\tilde M}^{*2} }
\newcommand{\est}{ {\tilde E}^* }
\newcommand{\bfgamma}{\mbox{\boldmath$\gamma$\unboldmath}}
\newcommand{\pabs}{|{\bf p}|}
\newcommand{\pabscm}{|{\bf p}_{c.m.}|}
\newcommand{\thetacm}{\theta_{c.m.}}
\newcommand{\qabs}{|{\bf q}|}
\begin{frontmatter}
\title{Covariant representations of the relativistic Brueckner T-matrix 
and the nuclear matter problem}
\author{T. Gross-Boelting,}
\author{C. Fuchs}
\author{and Amand Faessler}
\address{Institut f\"ur Theoretische Physik, 
Universit\"at T\"ubingen, D-72076 T\"ubingen, Germany}
\begin{abstract}
We investigate nuclear matter properties in the relativistic Brueckner
approach. The in-medium on-shell T-matrix
is represented covariantly by five Lorentz invariant amplitudes from which we 
deduce directly the nucleon self-energy. 
We discuss the ambiguities of this approach and
the failure of previously used covariant representations 
in reproducing the nucleon self-energies on the Hartree-Fock level. 
To enforce correct Hartree-Fock results we develop a subtraction scheme 
which treats the bare nucleon-nucleon potential exactly in accordance to the 
different types of meson exchanges. 
For the remaining ladder kernel, which contains the higher order correlations,
we employ then two different covariant 
representations in order to study the uncertainty inherent in the approach. 
The nuclear matter bulk properties are only 
slightly sensitive on the explicit representation used for the kernel. 
However, we obtain new Coester lines for the various Bonn potentials 
which are shifted towards the empirical region of saturation. In addition the 
nuclear equation-of-state turns out to be significantly softer in the new 
approach.
\end{abstract}
\end{frontmatter}
\section{Introduction}
The investigation of nuclear matter properties within the
relativistic Dirac-Brueckner-Hartree-Fock (DBHF) approach 
\cite{horowitz87,terhaar87,brockmann90,sehn97,fuchs98,dejong98} 
remains a fundamental topic in theoretical nuclear structure studies. 
Compared to non-relativistic approaches the relativistic DBHF treatment 
turned out to be a major step forward in the explanation of the saturation
mechanism of nuclear matter. The saturation points obtained for 
non-relativistic calculations, throughout all possible choices of different
nucleon-nucleon interactions, are located on the so called 'Coester line'
\cite{coester70} which does not meet the empirical saturation region. 
Using modern nucleon-nucleon interactions of the one-boson exchange type 
\cite{erkelenz74,holinde81} 
the relativistic calculations also reveal such Coester lines which are, however, 
significantly shifted towards the empirical region \cite{brockmann90}. 

On the other hand, many 
details of the relativistic theory are still not fully resolved. 
In particular, the precise form of the nucleon self-energy, i.e. 
the magnitude and the momentum dependence of the scalar and vector 
self-energy components are a question of current debate 
\cite{brockmann90,sehn97,fuchs98,nuppenau89,lee97}. 
Since the self-energy describes the dressing 
of the particles inside the medium and thus determines the relativistic 
mean field this fact states a severe problem. Different 
techniques to handle the DBHF problem can lead to significantly different 
results \cite{brockmann90,sehn97,fuchs98,dejong98,lee97}. 
In a recent work \cite{fuchs98} we found that 
the momentum dependence of the nucleon 
self-energy is dominated by the one-pion exchange contribution which 
accounts for the nuclear tensor force. 

Unfortunately, the treatment of the $\pi$NN vertex and the corresponding
self-energy contributions is closely 
connected to a severe ambiguity in the T-matrix 
representation \cite{fuchs98,nuppenau89}. 
The DBHF approach starts from a realistic nucleon-nucleon potential of the 
one-boson exchange type, i.e. the Bonn potentials \cite{machleidt86}.
As for the free two-body scattering problem, anti-particle states 
are neglected, and thus one works exclusively with positive energy states. 
Hence, a direct determination of the nucleon self-energy operator is 
not possible since not all matrix elements of this operator are known.
Horowitz and Serot have therefore developed a projection technique to 
determine the scalar and vector self-energy components 
from the in-medium T-matrix \cite{horowitz87}. 
In this approach the T-matrix is represented covariantly by 
Dirac operators and Lorentz invariant amplitudes where the latter are 
determined from the positive-energy on-shell T-matrix elements. 

The whole problem arises from the fact mentioned above, namely that one does not
include negative energy states and therefore neglects the excitation of 
anti-nucleons. The inclusion of negative energy excitations with 4 states for 
each spinor yields $4^4=256$ types of two-body matrix elements concerning their
spinor structure. Symmetry arguments reduce this to 44 for on-shell particles.
\cite{tjon85b}. If one takes now only positive energy solutions into account
this reduces to $2^4=16$ two-body matrix elements. Considering in addition only
on-shell matrix elements the number of independent matrix elements can be further
reduced by symmetry arguments down to 5. Thus, all on-shell two-body matrix elements
can be expanded into five Lorentz invariants. But these five invariants are not 
unique since the Dirac matrices involve always also negative energy states and thus
a decomposition of the one-body nucleon-nucleon potential into a Lorentz scalar and
a Lorentz vector contribution depends on the choice of these five Lorentz invariants
mentioned above. The best choice would be to separate completely the negative energy
Dirac states. But since this is not possible, there is not a unique but only an
'optimal choice'. The topic of this paper is the form of this 'optimal choice' of the
five invariants.

Thus, as discussed in \cite{fuchs98}, various covariant 
representations of the T-matrix exist which all reproduce identically the on-shell 
T-matrix elements but lead to rather different results for the nucleon 
self-energy in the present formalism. This ambiguity was found to arise substantially from 
the treatment of the pion exchange part, in particular by the way 
how to take the pseudo-vector nature 
of the $\pi$NN vertex into account. In realistic nucleon-nucleon
potentials the pion is usually described by 
a pseudo-vector coupling \cite{machleidt86}. 
A pseudo-vector $\pi$NN vertex is also predicted by the non-linear $\sigma$-model 
based on chiral symmetry \cite{weinberg68}. 
Furthermore, a pseudo-vector $\pi$NN vertex has the advantage 
that it is consistent with neglection of 
negative energy states while a pseudo-scalar coupling connects very strongly positive and
negative energy states. 

Following this argumentation in several 
works \cite{terhaar87,sehn97,nuppenau89,terhaar87b}
the so called 'pseudo-vector choice' to the T-matrix representation was 
applied. This means to simply replace the pseudo-scalar by a pseudo-vector 
covariant while keeping the corresponding amplitudes unchanged. 
However, this procedure is not well defined since equivalent $ps$ representations 
exist which lead to non-equivalent $pv$ representations.
A strong momentum dependence of the nucleon 
self-energy emerges, e.g., in the 'conventional' $pv$ representation as it 
was applied by Sehn et al. \cite{sehn97}. In Ref. \cite{fuchs98} we 
addressed the problem with the above ambiguity, in particular with respect 
to the determination of the nuclear self-energy components. 
There we also proposed a 'complete' $pv$ representation of the T-matrix 
which results in a much weaker momentum 
dependence of the nucleon self-energy as found in \cite{sehn97}. 
The latter representation is not
only more consistent with the approximation scheme of the current DBHF 
approach but also works correctly at the Hartree-Fock 
level using the pseudo-vector pion exchange. 
This minimal requirement, namely that the complete procedure 
is able to reproduce the correct Hartree-Fock results, 
was never verified for other representations and indeed, the 'conventional' 
$pv$ representation fails in this respect.
Hence, a strong momentum dependence observed for the nucleon
self-energy \cite{sehn97} appears to be the artifact of a 
misrepresentation of the pion exchange potential in the
previously used 'conventional' $pv$ representations 
\cite{terhaar87,sehn97,terhaar87b}.

In this paper we want to go beyond the investigation of 
Ref. \cite{fuchs98}. There we tried to determine the 
range of uncertainty inherent in the present 
Brueckner approach by the consideration of the two limiting 
cases, namely the pseudo-scalar and the complete pseudo-vector 
representation of the T-matrix. One has, however, some additional 
'leading order' information on the Lorentz structure of the T-matrix, 
which is given by the Born part, i.e. the 
bare nucleon-nucleon potential. If we want to reproduce 
the analytically known Hartree-Fock results 
\cite{fritz94} for the complete set of the six non-strange 
mesons used in the Bonn potential, 
only a mixed '$ps+pv$' representation can be successful 
where the different parts of the bare potential are represented separately 
either as $ps$ or $pv$. It is clear that such a mixed representation is not 
feasible at the DBHF level since we can not disentangle the different meson 
contributions from the full in-medium interaction. To proceed, however,  
we suggest a subtraction method to the in-medium T-matrix which 
means to subtract the bare interaction, i.e. the leading order, 
from the full T-matrix and to treat the
bare part - the Hartree-Fock level - in the mixed '$ps+pv$' representation.
The remaining ladder kernel or 'subtracted T-matrix', as we will call it
in the following, i.e. the sum of all higher order exchange graphs, is
then represented in different ways, either as pure $ps$ or 'complete' 
$pv$. Applying these two representation schemes we are now able 
to reduce the range of uncertainty concerning the T-matrix 
representation to a minimum. The remaining uncertainty 
is inherent in the current DBHF approach and can not be removed 
by standard methods. The deviations in the final results are, however, 
small and -- similar to the treatment of Ref. \cite{fuchs98} -- the momentum 
dependence of the self-energy is found to be rather weak. 
Furthermore we obtain new 'Coester lines' for Bonn A, B, C with, 
compared to previous works \cite{brockmann90,sehn97}, improved saturation 
properties. Most remarkably is a strong softening of the 
nuclear equation-of-states.

The paper is now organized as follows: 
In section 2 we briefly review the Dirac-Brueckner Hartree-Fock approach and 
present some details on the calculation of the in-medium on-shell 
T-matrix elements.
In section 3 we introduce the projection technique and discuss the 
different covariant representations used for the on-shell T-matrix. 
The subtraction method is developed at the end of this section.
The nucleon self-energy in the medium and the nuclear matter 
bulk properties are then presented elaborately in section 4. 
At the end we summarize and conclude our work.
\section{The relativistic Brueckner approach}
\subsection{The coupled set of equations}
In the relativistic Brueckner approach the nucleon 
inside the nuclear medium is viewed as a dressed particle in consequence
of its two-body interaction with the surrounding nucleons. 
The in-medium interaction of the nucleons is treated in the ladder
approximation of the relativistic Bethe-Salpeter equation
\beq
T= V + i \int  VQGGT
\quad ,
\label{BSeq}
\eeq
where $T$ denotes the T-matrix. $V$ is the bare nucleon-nucleon interaction. 
The intermediate off-shell nucleons in the 
scattering equation are described by a two-particle propagator $iGG$.
As usually done, we  replace this propagator by an effective propagator \cite{brown76},
here the Thompson propagator \cite{thompson70} which allows only positive 
energy nucleons in the intermediate scattering states. In addition this 
propagator fixes also the off-shell behavior of the nucleons. This reduces the 
four-dimensional Bethe-Salpeter equation to a three-dimensional integral
equation. The Pauli operator $Q$ in the Thompson equation accounts for the 
influence of the medium by the Pauli-principle and projects the 
intermediate scattering states out of the Fermi sea. 

The Green's function $G$ which describes the propagation of dressed 
nucleons in the medium fulfills the Dyson equation
\beq
G=G_0+G_0\Sigma G 
\quad .
\label{Dysoneq}
\eeq 
$G_0$ denotes the free nucleon propagator while the influence of the 
surrounding nucleons is expressed by the nucleon self-energy $\Sigma$. 
In Brueckner theory this self-energy is determined by summing up the 
interaction with all the nucleons inside the Fermi sea 
\beq
\Sigma = -i \int\limits_{F} (Tr[G T] - GT )
\quad .
\label{HFselfeq1}
\eeq
The Hartree-Fock form of the self-energy integral is necessary if we
use an 'unphysical' T-matrix, as done in \cite{horowitz87}. 
However, since we will entirely work with anti-symmetrized two-nucleon states,  
our T-matrix is 'physical' and contains implicitly 'direct' and 'exchange' 
contributions. Hence, the Hartree form 
\beq
\Sigma = -i \int\limits_{F} Tr[G T] 
\label{HFselfeq2}
\eeq
of the self-energy integral is sufficient in our case. 
We will come back to this point in more detail in section 3. 
Since the three equations (\ref{BSeq}), (\ref{Dysoneq}) and 
(\ref{HFselfeq2}) are strongly coupled, one has to 
solve this set of equations self-consistently. 

The Dirac structure of the self-energy in isospin saturated nuclear matter 
follows from translational and rotational invariance, parity conservation and 
time reversal invariance \cite{serot86}. 
In the nuclear matter rest frame the self-energy has the simple form 
\beq
\Sigma(k,\kf)= \Sigs (k,\kf) -\gamma_0 \, \Sigo (k,\kf) + 
{\bfgamma}\cdot {\bf k} \,\Sigv (k,\kf) 
\quad ,
\label{self1}
\eeq
with $k_\mu$ being the nucleon four-momentum. 
The self-energy components depend as Lorentz scalar functions 
on the Lorentz invariants $k^2$, $k\cdot j$ and $j^2$, where 
$j_\mu$ denotes the baryon current. In the nuclear matter rest frame
this current is identical to $j_\mu=\rho\delta_{\mu 0}$, with $\rho$ 
being the nuclear matter density. 
Hence, the Lorentz invariants can be expressed in terms of 
$k_0, |{\bf k}|$ and $\kf$, where $\kf$ denotes the Fermi momentum, related
to the density via $\rho=2\kf^3/(3\pi^2)$. 
By taking the traces in Dirac space as \cite{horowitz87,sehn97}
\beq
\Sigs = \frac{1}{4} tr \left[ \Sigma \right] \quad ,\quad 
\Sigo = \frac{-1}{4} tr \left[ \gamma_0 \, \Sigma \right]
\quad , \quad 
\Sigv =  \frac{-1}{4|{\bf k}|^2 } 
tr \left[{\bfgamma}\cdot {\bf k} \, \Sigma \right] 
\label{trace}
\eeq
one can calculate the different Lorentz components of the self-energy from
the self-energy operator (\ref{HFselfeq2}). 

The presence of the medium leads to effective masses and effective momenta 
\beq
m^*(k,\kf)= M + Re \Sigs (k,\kf) \quad , 
\quad k_\mu^* = k_\mu + Re  \Sigma_\mu(k,\kf)
\eeq
of the nucleons. Above the Fermi surface the self-energy is 
generally complex due to possible 
decay of particle states into hole states within the Fermi sea. 
To simplify the self-consistency scheme we neglect this decay process 
and work in the so called 'quasi-particle approximation'. 
Since we only deal with the real part of the self-energy we omit this in the 
notation from now on.

Defining reduced effective quantities \cite{sehn97,fuchs98} 
\beq
{{\tilde k}^*}_\mu = k^*_\mu / \left( 1+\Sigv(k,\kf)\right) \, , \,
{\tilde m}^*(k,\kf) = m^*(k,\kf)/ \left( 1+\Sigv(k,\kf)\right)
\label{redquantity}
\quad ,
\eeq
the Dirac equation in the nuclear matter rest frame can be rewritten as
\beq
 [ \gamma_\mu {\tilde k}^{*^\mu} - {\tilde m}^*(k,\kf)] u(k,\kf)=0
\quad 
\label{dirac2}
\eeq
which resembles a quasi-free Dirac equation for dressed
nucleons. In general the reduced effective mass is density but also 
momentum dependent. 
To simplify the calculation, however, we fix the effective mass of the nucleon 
in the nuclear matter rest frame at the reference momentum $|{\bf k}|=k_F$. 
In this 'reference spectrum approximation' \cite{bethe63} the reduced 
effective mass $\tilde{m}^*_F = \tilde{m}^*(|{\bf k}|=\kf)$ works as a 
self-consistency parameter in the current DBHF approach. 
All equations are iterated until $\tilde{m}^*_F$ is converged to a fixed 
value. The 'reference spectrum approximation' implies that the self-energy 
itself is rather weakly momentum dependent. 
At the end of the calculation one has to verify the consistency of the 
assumption $\Sigma(k)\approx \Sigma(|{\bf k}|=\kf)$ 
with the outcome of the iteration procedure.

Utilizing the different approximations discussed above the positive-energy 
in-medium nucleon spinor are given as
\beq
u_\lambda (k,\kf)= \sqrt{ { {\tilde E}^*({\bf k})+ {\tilde m}^*_F}\over
{2{\tilde m}^*_F}} 
\left( 
\begin{array}{c} 1 \\ 
{2\lambda |{\bf k}|}\over{{\tilde E}^*({\bf k})+ {\tilde m}^*_F}
\end{array}
\right)
\chi_\lambda
\quad ,
\eeq
where ${\tilde E}^*({\bf k})=\sqrt{{\bf k}^2+{\tilde m}^{*2}_F}$.
$\chi_\lambda$ above denotes a two-component Pauli spinor with 
$\lambda=\pm {1\over 2}$ and the normalization of the Dirac spinor is 
$\bar{u}_\lambda(k,\kf) u_\lambda(k,\kf)=1$. Since the in-medium nucleon 
spinor contains the reduced effective mass the matrix elements of the 
bare nucleon-nucleon interaction are density dependent. 
This density effect does not appear in non-relativistic Brueckner
calculations. It is believed that it is the main reason 
for the success of the DBHF approach in describing the saturation of nuclear 
matter \cite{anastasio83}.
\subsection{The in-medium T-matrix}
We apply the relativistic Thompson equation \cite{thompson70} to solve the 
scattering problem of two nucleons in the nuclear medium. 
In the two-particle center of mass (c.m.) 
frame - the natural frame for studying the two-particle 
scattering process - this Thompson equation can be written as 
\cite{terhaar87,sehn97}
\beqa
T({\bf p},{\bf q},x)|_{c.m.} &=&  V({\bf p},{\bf q}) 
\label{Tmateq}\\
&+& 
\int {d^3{\bf k}\over {(2\pi)^3}}
{\rm V}({\bf p},{\bf k})
{{\tilde m}^{*2}_F\over{\tilde{E}^{*2}({\bf k})}}
{{Q({\bf k},x)}\over{2{\tilde{E}}^*({\bf q})-2{\tilde{E}}^*({\bf k})
+i\epsilon}}
T({\bf k},{\bf q},x) 
\nonumber
\quad ,
\label{thompsoneq}
\eeqa
where ${\bf q}=({\bf q}_1 - {\bf q}_2)/2$ is the relative three-momentum 
of the initial state while $\bf k$ and $\bf p$ are the relative 
three-momenta of the intermediate and final states, respectively. 
The total four-momentum of the two-nucleon
system is $\tilde{P}^*=\tilde{q}^*_1+\tilde{q}^*_2$, which in the 
c.m. frame becomes $\tilde{P}^*=(\tilde{P}^*_0,\bf 0)$. 
$\tilde{P}^*_0=\sqrt{\tilde{s}^*}=2{\tilde{E}}^*({\bf q})
=2\sqrt{{\bf q}^2+\tilde{m}^{*2}_F}$ is the starting energy in
(\ref{Tmateq}). If ${\bf q}_1$ and ${\bf q}_2$ are 
nuclear matter rest frame momenta of the nucleons in the initial state, 
the boost-velocity $\bf u$ into the c.m. frame is given by 
\beq
{\bf u} = {\bf P}/ \sqrt{\tilde{s}^{*}+{\bf P}^2}
\label{boost}
\quad ,
\eeq
with the total three-momentum and the invariant mass 
${\bf P} = {\bf q}_1+{\bf q}_2$ and $\tilde{s}^*=(\tilde{E}^*({\bf q}_1)
+\tilde{E}^*({\bf q}_2))^2-{\bf P}^2$, respectively.
In Eq. (\ref{thompsoneq}) x denotes the set of 
additional parameters $x=\{\kf, {\tilde m}^*_F,|{\bf u}|\}$ 
on which the T-matrix depends.

Applying standard techniques as explained in detail by Erkelenz \cite{erkelenz74}
we solve the Thompson equation in the c.m. frame and calculate the 
plane-wave helicity matrix elements of the T-matrix. 
To determine the self-energy only positive-energy 
T-matrix elements at on-shell points $|{\bf p}|= |{\bf q}|$ are necessary 
since in (\ref{HFselfeq2}) we use instead of the full nucleon propagator 
the Dirac propagator \cite{horowitz87,serot86}
\beq
G^D(q)=[\gamma_\mu \tilde{q}^{*\mu}+\tilde{m}^*_F]2\pi i \delta (\tilde{q}^{*2}
- \tilde{m}^{*2}_F)\Theta(\tilde{q}^*_0)\Theta(\kf-|{\bf q}|)
\label{diracprop}
\quad .
\eeq
Here q denotes the nuclear matter rest frame momentum of the nucleon in
the Fermi sea. This momentum is on-mass shell, therefore only 
elastic scattering amplitudes, i.e. on-shell T-matrix elements, contribute 
to the nucleon self-energy. 
Due to the $\Theta$-functions in the propagator only positive energy nucleons
are allowed in the intermediate scattering state. Hence the subspace of 
negative energy states is omitted in the current Brueckner 
approach. In this way we avoid the delicate problem of infinities in the
theory which would generally appear if we would include contributions 
from negative energy nucleons in the Dirac sea 
\cite{horowitz87,dejong98,poschenrieder88}. 

In the on-shell case parity and spin conservation demand that 
only five of sixteen possible positive-energy helicity matrix elements 
are linearly independent \cite{brown76}. The five matrix elements are determined 
explicitly via the $|{\rm JMLS}>$ scheme. 
After a partial-wave projection onto the $|{\rm JMLS}>$ states, 
the Thompson equation reduces to a partially decoupled set of 
one-dimensional integral equations over the relative momentum $|{\bf k}|$. 
To accomplish such a reduction an angle-averaged Pauli-operator $\overline Q$ 
is used instead of the full Pauli-operator $Q$. 
Since the Fermi sphere is deformed to a Fermi ellipsoid 
in the two-nucleon c.m. frame, $\overline Q$ has to be evaluated for such 
a Fermi ellipsoid. The explicit expression for $\overline Q$ can be found in
Refs. \cite{terhaar87,sehn97,sehn98}. 
Finally, the integral equations are solved numerically by the
matrix inversion technique of Haftel and Tabakin \cite{haftel70}.

Since the two-nucleon states are two-fermion states, we actually have to 
evaluate the fully anti-symmetrized matrix elements. Only these matrix
elements are physically meaningful. Anti-symmetrization is achieved by 
restoring the total isospin of the two-nucleon system $({\rm I}=0,1)$ with 
the help of the standard selection rule 
\beq
(-1)^{\rm L+S+I}=-1
\label{selection}
\quad .
\eeq 
The five on-shell plane-wave helicity matrix elements 
for definite isospin are finally 
calculated from the five partial-wave helicity matrix elements obtained in the
$|{\rm JMLS}>$ scheme by summing over the total angular momentum J as 
\beqa
<{\bf p} \lambda_1^{'} \lambda_2^{'}| T^{\rm I}(x)|
{\bf q} \lambda_1 \lambda_2>
= \sum\limits_{\rm J} \left( \frac{2{\rm J}+1}{4\pi}\right) 
d^J_{\lambda^{'} \lambda}(\theta)
<{\bf p} \lambda_1^{'} \lambda_2^{'}| T^{\rm J,I}(x)|
{\bf q} \lambda_1 \lambda_2> .
\nonumber \\
\label{tmatel1}
\eeqa
$\theta$ is the scattering angle between $\bf q$ and $\bf p$, with
$\pabs=\qabs$, while $\lambda=\lambda_1-\lambda_2$ and 
$\lambda^{'}=\lambda_1^{'}-\lambda_2^{'}$.
The reduced rotation matrices $d^{\rm J}_{\lambda^{'} \lambda}(\theta)$ 
are those of Rose \cite{rose57}.
\section{Covariant representations and the nucleon self-energy}
\subsection{Pseudo-scalar representation}
To use the trace formulas, Eqs. (\ref{trace}), for the calculation of the 
nucleon self-energy components one has to represent the T-matrix in the 
nuclear matter rest frame. Since we determine the T-matrix
elements in the two-particle c.m. frame a representation with  
covariant operators and Lorentz invariant amplitudes 
in Dirac space is the most convenient way to Lorentz-transform the T-matrix 
from one frame into another \cite{horowitz87}. 
A set of five linearly independent covariants is 
sufficient for such a T-matrix representation because on-shell only 
five helicity matrix elements appear as solution of the Thompson equation. 
A linearly independent although not unique set of five covariants 
is given by the Fermi covariants 
\beqa
\!\!\!\!\!{\rm S} = 1\otimes 1 ,
{\rm V} =  \gamma^{\mu}\otimes \gamma_{\mu},
{\rm T} = \sigma^{\mu\nu}\otimes\sigma_{\mu\nu}, 
{\rm A} =  \gamma_5 \gamma^{\mu}\otimes \gamma_5 \gamma_{\mu},
{\rm P} = \gamma_5 \otimes \gamma_5 .
\eeqa
Using this special set - the so called 'pseudo-scalar choice' - 
the on-shell T-matrix for definite isospin I can be represented 
covariantly as \cite{horowitz87}
\beqa
\hspace{1cm}
T^{\rm I}(\pabs,\theta,x)&=& 
 F_{\rm S}^{\rm I}(\pabs,\theta,x){\rm S}
+F_{\rm V}^{\rm I}(\pabs,\theta,x){\rm V}
+F_{\rm T}^{\rm I}(\pabs,\theta,x){\rm T}
\nonumber \\
&+&F_{\rm A}^{\rm I}(\pabs,\theta,x){\rm A}
+F_{\rm P}^{\rm I}(\pabs,\theta,x){\rm P}
\label{tmatrep1}
\quad .
\eeqa
Here ${\bf p}$ and $\theta$ denote the relative three-momentum and the
scattering angle between the scattered nucleons in the c.m. frame, 
respectively. In addition, the five Lorentz invariant 
amplitudes $F_i^{\rm I}(\pabs,\theta,x)$ with $i=\{{\rm S,V,T,A,P}\}$
depend also on $x=\{\kf, {\tilde m}^*_F,|{\bf u}|\}$. 
We want to stress that the dependence of the amplitudes on $\pabs$, $\theta$ and $x$
can be re-expressed in terms of Lorentz scalar quantities as explained in 
detail in \cite{horowitz87}. Hence, the representation (\ref{tmatrep1}) 
is indeed fully covariant, only the calculation of the 
Lorentz invariant amplitudes is most easily performed 
in the two-particle c.m. frame.

We determine the Lorentz invariant amplitudes 
$F_i^{\rm I}(\pabs,\theta,x)$ by taking 
plane-wave helicity matrix elements of (\ref{tmatrep1}) in the c.m. frame. 
Using as abbreviation for the Fermi covariants the operators 
$\Gamma_i$, with $\Gamma_i=\{{\rm S,V,T,A,P}\}$, the on-shell matrix elements 
($|{\bf p}|=|{\bf q}|$), i.e. the solution of Eq. (\ref{tmatel1}), and the
invariant amplitudes are related by 
\beq
<{\bf p}\lambda_1^{'}\lambda_2^{'}|T^{\rm I}(x)|
{\bf q}\lambda_1\lambda_2>
=\sum\limits_i<{\bf p}\lambda_1^{'}\lambda_2^{'}|\Gamma_i|
{\bf q}\lambda_1\lambda_2> F_i^{\rm I}(\pabs,\theta,x)
\quad .
\label{Tmatinv}
\eeq
Since the Fermi covariants are linearly independent, equation 
(\ref{Tmatinv}) can be inverted 
to determine the unknown amplitudes $F_i^{\rm I}$. 
The details of this inversion, especially the treatment of 
the kinematical singularities at $\theta=0$ and $\theta=\pi$, which appear
likewise in the matrix elements 
$d^{\rm J}_{\lambda^{'} \lambda}(\theta)$, Eq. (\ref{tmatel1}) 
and $<{\bf p}\lambda_1^{'}\lambda_2^{'}|\Gamma_i|{\bf q}\lambda_1\lambda_2>$, 
are explained in \cite{horowitz87,tjon85a}.

Using physical plane-wave helicity matrix elements in (\ref{Tmatinv}) 
the Lorentz invariant amplitudes $F_i^{\rm I}(\pabs,\theta,x)$ fulfill 
a specific anti-symmetry relation which is given by the 
Fierz transformation 
\beq
F_i^{\rm I}(\pabs,\pi-\theta,x)=
(-1)^{\rm I}F_j^{\rm I}(\pabs,\theta,x){\cal F}_{ji}
\quad ,
\label{antisymeq1}
\eeq
where the Fierz matrix ${\cal F}$ is the matrix given in Eq. (\ref{fierz}).
Due to this anti-symmetry relation the 'direct' representation 
(\ref{tmatrep1}) for the T-matrix is sufficient and one can calculate 
the nucleon self-energy via the Hartree integral (\ref{HFselfeq2}). 
Using explicitly momenta and Dirac indices, the self-energy 
reads
\beq
\Sigma_{\alpha\beta}(k,\kf)= -i \int {{d^4 q}\over{(2\pi)^4}}
G^D_{\tau\sigma}(q)
T(|{\bf p}|,0,x)_{\alpha\sigma,\beta\tau}
\quad .
\label{selfint}
\eeq
Here k and q denote the incoming and outgoing momenta of the two 
elastically scattered nucleons in the nuclear matter rest frame. 
The total energy of the two nucleons is   
$\tilde{s}^*=(\tilde{E}^*({\bf k})+\tilde{E}^*({\bf q}))^2-{\bf P}^2$ 
and the total three-momentum, which defines the boost (\ref{boost}) 
into the two-particle c.m. frame, is ${\bf P}={\bf k}+{\bf q}$.
The relative momentum in the c.m. frame for which we determine the T-matrix 
is given by $|{\bf p}|=\sqrt{\tilde{s}^*/4-\tilde{m}^{*2}_F}$. 
Since we only calculate the Hartree integral, the scattering
angle is fixed to $\theta=0$.

Applying the covariant representation (\ref{tmatrep1}) for the on-shell
T-matrix the nucleon self-energy in isospin saturated nuclear matter 
is evaluated to be \cite{sehn97}
\beq
\Sigma_{\alpha\beta}(k,\kf)= \int {{d^3{\bf q}}\over {(2\pi)^3}}
{{\theta(\kf-|{\bf q}|)}\over {\tilde{E}^*({\bf q})}}
\left[\tilde{m}^*_F 1_{\alpha\beta}F_{\rm S}
+\not{\tilde q}^*_{\alpha\beta}F_{\rm V}\right]
\quad ,
\label{self1}
\eeq
where the isospin averaged amplitudes are defined as
\beq
F_i(|{\bf p}|,0,x):=
{1\over 2}\left[ F_i^{{\rm I}=0}(|{\bf p}|,0,x)
+3 F_i^{{\rm I}=1}(|{\bf p}|,0,x)\right]
\quad .
\eeq
In the self-energy integral (\ref{self1}) only the anti-symmetrized scalar and 
vector amplitudes $F_S$ and $F_V$ for scattering angle $\theta=0$ (Hartree) contribute.
This is not true if we use 'unphysical' helicity matrix 
elements as done by Horowitz and Serot \cite{horowitz87}. 
If we neglect the selection rule (\ref{selection}),
we have to determine explicitly the Hartree and the Fock contribution 
to the self-energy via
\beq
\Sigma_{\alpha\beta}(k,\kf)= -i \int {{d^4 q}\over{(2\pi)^4}}
G^D_{\tau\sigma}(q) 
[T(|{\bf p}|,0,x)_{\alpha\sigma,\beta\tau}
-T(|{\bf p}|,\pi,x)_{\alpha\sigma,\tau\beta}]
\quad .
\label{selfint2}
\eeq
An equivalent procedure is to use the Hartree form (\ref{selfint})
but with a T-matrix representation which explicitly contains
'direct' and 'exchange' terms. This is done by calculating  
'unphysical' Lorentz invariant amplitudes 
${\bar F}_i^{\rm I}$ for scattering angles $\theta$ and $\pi - \theta$,  
using Eq. (\ref{Tmatinv}) with non-antisymmetrized plane-wave helicity 
matrix elements. Then one defines interchanged Fermi covariants as 
\cite{tjon85a}
\beq
\tilde{{\rm S}}=\tilde{{\rm S}}{\rm S}\quad , \quad
\tilde{{\rm V}}=\tilde{{\rm S}}{\rm V}\quad , \quad
\tilde{{\rm T}}=\tilde{\rm S}{\rm T}\quad , \quad
\tilde{\rm A}=\tilde{\rm S}{\rm A} \quad ,\quad 
\tilde{{\rm P}}=\tilde{{\rm S}}{\rm P}
\label{excovariants}\quad ,
\eeq
where the operator $\tilde{\rm S}$, being the interchange covariant of 
{\rm S}, exchanges the Dirac indices of particle
1 and 2, i.e. $\tilde{\rm S}u(1)_\sigma u(2)_\tau = u(1)_\tau u(2)_\sigma$.
The on-shell T-matrix is finally decomposed to 
\beq
T^{{\rm I}}(\pabs,\theta,x)=T^{{\rm I},D}(\pabs,\theta,x)
-T^{{\rm I},X}(\pabs,\theta,x)
\quad ,
\label{tmatrep2}
\eeq
where the 'direct' part of the T-matrix is defined as 
\beqa
T^{{\rm I},D}(\pabs,\theta,x)&=& \left[ 
\bar{F}_{\rm S}^{\rm I}(\pabs,\theta,x){\rm S}+
\bar{F}_{\rm V}^{\rm I}(\pabs,\theta,x){\rm V}+
\bar{F}_{\rm T}^{\rm I}(\pabs,\theta,x){\rm T} \right.
\nonumber \\
&+&\left. \bar{F}_{\rm A}^{\rm I}(\pabs,\theta,x){\rm A}+
\bar{F}_{\rm P}^{\rm I}(\pabs,\theta,x){\rm P} \right] 
\quad ,
\eeqa
while the 'exchange' part is given as 
\beqa
&&T^{{\rm I},X}(\pabs,\theta,x)= (-1)^{\rm I+1}\left[ 
\bar{F}_{\rm S}^{\rm I}(\pabs,\pi-\theta,x)\tilde{{\rm S}}+
\bar{F}_{\rm V}^{\rm I}(\pabs,\pi-\theta,x)\tilde{{\rm V}}
\right. \nonumber \\
&&\left. \,\,\,\,\,\,\,\,
+\bar{F}_{\rm T}^{\rm I}(\pabs,\pi-\theta,x)\tilde{{\rm T}} 
+\bar{F}_{\rm A}^{\rm I}(\pabs,\pi-\theta,x)\tilde{{\rm A}}
+\bar{F}_{\rm P}^{\rm I}(\pabs,\pi-\theta,x)\tilde{{\rm P}} \right]
.
\eeqa
Since the interchanged and original Fermi covariants are related by the 
Fierz transformation ${\cal F}$ \cite{tjon85a} via  
\beqa
\left( 
\begin{array}{c} \tilde{{\rm S}} \\ \tilde{\rm V} \\ \tilde{\rm T} \\ 
\tilde{\rm A} \\ \tilde{\rm P} \end{array} 
\right)
= {1\over 4}
\left( 
\small{ 
\begin{array}{ccccc} 
 1  &   1 & {1\over 2} & -1   &  1  \\
 4  &  -2 &  0         & -2  &  -4  \\
 12 &   0 &  -2        & 0  &  12   \\
 -4 &  -2 &  0         & -2   &  4 \\
 1  &  -1 & {1\over 2} & 1   &  1 
\end{array}} 
\label{fierz}
\right)
\left( 
\begin{array}{c} {\rm S} \\ {\rm V} \\ {\rm T} \\ {\rm A} \\ {\rm P} 
\end{array} 
\right) 
\quad ,
\label{transform}
\eeqa
the 'exchange' contribution of the T-matrix is identical to 
\beq
T^{{\rm I,X}}(\pabs,\theta,x)=(-1)^{\rm I+1}
\sum_{ij}\bar{F}_i^{\rm I}(\pabs,\pi-\theta,x){\cal F}_{ij}\Gamma_j
\quad .
\eeq
Above only the Fermi covariants $\Gamma_i$ appear and one can rewrite 
the T-matrix representation (\ref{tmatrep2}) 
in terms of a 'direct' representation as in (\ref{tmatrep1}). 
The fully anti-symmetrized Lorentz invariant amplitudes are therefore related to
the non-antisymmetrized amplitudes via the identity
\beq
F_i^{\rm I}((\pabs,\theta,x)= \bar{F}_i^{\rm I}((\pabs,\theta,x)-
(-1)^{\rm I+1} \bar{F}_j^{\rm I}((\pabs,\pi-\theta,x){\cal F}_{ji}
\label{antisymeq2}
\quad .
\eeq
These anti-symmetrized amplitudes respect the relation (\ref{antisymeq1}) 
and are identical to the amplitudes which we obtain when we use 
'physical' plane-wave helicity matrix elements in (\ref{Tmatinv}) from the
beginning. Nevertheless when employing the T-matrix representation 
(\ref{tmatrep2}) with the non-antisymmetrized amplitudes one obtains 
for the self-energy the expression given in Ref. \cite{horowitz87}
\beqa
\Sigma_{\alpha\beta}(k,\kf)&=& \int {{d^3{\bf q}}\over {(2\pi)^3}}
{{\theta(\kf-|{\bf q}|)}\over {4\tilde{E}^*({\bf q})}}
\left[
\not{\tilde q}^*_{\alpha\beta}
\left(
4\bar{F}_{\rm V}^D-\bar{F}_{\rm S}^X+2\bar{F}_{\rm V}^X+2\bar{F}_{\rm A}^X
+\bar{F}_{\rm P}^X \right)
\right. \nonumber \\
&&\left. +\tilde{m}^*_F 1_{\alpha\beta} 
\left(
4\bar{F}_{\rm S}^D-\bar{F}_{\rm S}^X-4\bar{F}_{\rm V}^X-12\bar{F}_{\rm T}^X
+4\bar{F}_{\rm A}^X
-\bar{F}_{\rm P}^X\right) \right]
\label{self2}
\eeqa
where the isospin-averaged non-antisymmetrized amplitudes are defined as
\beqa
\hspace{2cm}
\bar{F}_i^D&:=&{1\over 2}\left[\bar{F}_i^{\rm I=0}(|{\bf p}|,0,x)
+3\bar{F}_i^{\rm I=1}(|{\bf p}|,0,x)\right] \nonumber \\
\bar{F}_i^X&:=&{1\over 2}\left[-\bar{F}_i^{\rm I=0}(|{\bf p}|,\pi,x)
+3\bar{F}_i^{\rm I=1}(|{\bf p}|,\pi,x)\right]
\quad .
\eeqa
Due to relation (\ref{antisymeq2}) the nucleon
self-energy defined via the integrals (\ref{self1}) or (\ref{self2}) 
is, of course, identical.

\begin{figure}[h]
\vspace{75mm}
\includegraphics{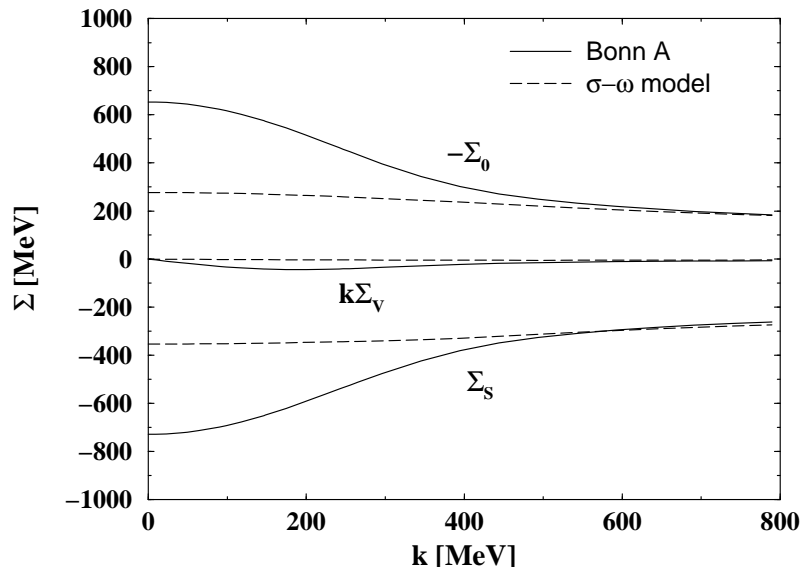}
\caption{\label{fig1} Momentum dependence of the DBHF nucleon self-energy
components in nuclear matter at $\kf=1.34 fm^{-1}$ using as bare 
nucleon-nucleon potential Bonn A (solid) and the $\sigma$-$\omega$ model 
potential (dashed). For the T-matrix the $ps$ representation 
(\ref{tmatrep1}) is applied.}
\end{figure}

\begin{figure}[h]
\vspace{75mm}
\includegraphics{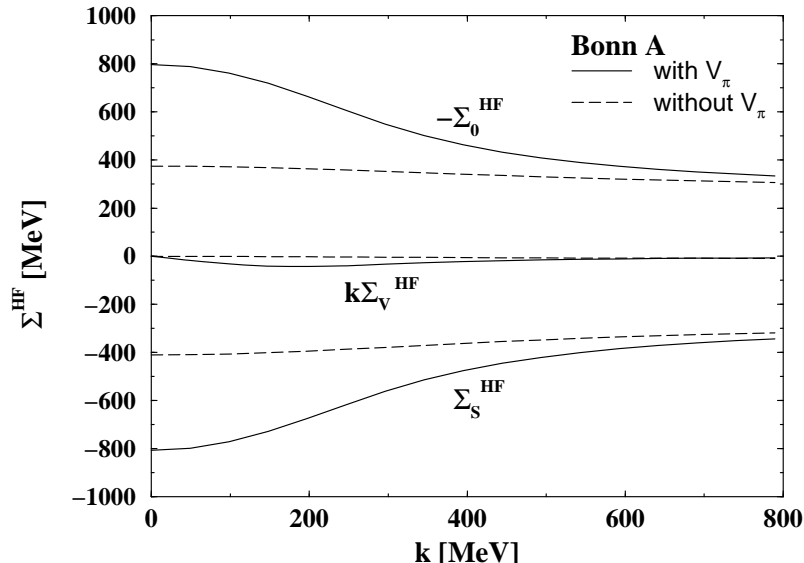}
\caption{\label{fig2} Momentum dependence of the Hartree-Fock 
nucleon self-energy components in nuclear matter at $\kf=1.34 fm^{-1}$.
The reduced effective mass of the nucleon is fixed at $\tilde{m}^*_F=500 MeV$.
As nucleon-nucleon interaction the Bonn A potential with (solid) and 
without (dashed) pion exchange part is used. For the bare interaction 
the $ps$ representation (\ref{tmatrep1}) is applied.}
\end{figure}

In Fig. \ref{fig1} we show the result of a self-consistent DBHF calculation for the 
nucleon self-energy in nuclear matter applying as representation for the
on-shell T-matrix the $ps$ representation (\ref{tmatrep1}).
As bare interaction we have used the Bonn A potential \cite{machleidt86} 
and, for comparison, the $\sigma$-$\omega$ model potential \cite{walecka74}
which was originally used by 
Horowitz and Serot \cite{horowitz87}. 
The density is chosen to be the empirical saturation density of nuclear matter 
with a Fermi momentum of $\kf=1.34 fm^{-1}$.
As already discussed in Ref. \cite{sehn97}, we see a pronounced 
momentum dependence of the nucleon self-energy 
components with the full Bonn A while in the case of the $\sigma$-$\omega$ 
model potential the dependence on the momentum is rather weak. 
A strong momentum dependence questions, of course, the validity of the 
'reference spectrum approximation' used in the present self-consistency 
scheme. Furthermore, such a strong momentum dependence leads to unphysical
results deep inside the Fermi sea since the effective mass drops to
values which are close to zero. Therefore in Ref. \cite{fuchs98}
the strong momentum dependence of the self-energy was studied in more detail 
and found to originate mainly from the one-pion exchange contribution to the
self-energy. To illustrate this aspect, in Fig. \ref{fig2} the result of a 
non-selfconsistent Hartree-Fock (HF) calculation are shown. 
The HF nucleon self-energy is defined via the integral
\beq
\Sigma^{HF} = -i \int\limits_{F} Tr[G^{D} V] 
\quad .
\label{HFselfeq3}
\eeq
As in the case of the full DBHF calculation we determine at first the matrix elements
of V and apply afterwards the $ps$ representation (\ref{tmatrep1}) for the
matrix elements. For better comparison to Fig. \ref{fig1} we have 
fixed the Fermi momentum again at $\kf=1.34 fm^{-1}$ while for the 
effective mass, necessary to calculate the propagator $G^D$ and the dressed  
potential V, we have taken a fixed value of $\tilde{m}^*_F=500 MeV$.
With the full Bonn potential the Hartree-Fock nucleon self-energy 
is similar to the result of the self-consistent DBHF calculation. 
On the other hand applying Bonn A without the pion exchange part, the 
Hartree-Fock nucleon self-energy is weakly momentum dependent, as previously
observed in the DBHF calculation with the $\sigma$-$\omega$ model 
potential. This shows that the pion exchange part within the 
Bonn potential 
is responsible for the strong momentum dependence in the nucleon self-energy.
That the non-selfconsistent Hartree-Fock 
nucleon self-energy is also similar to the DBHF 
nucleon self-energy in the case of the Bonn A potential 
indicates that the summation of the higher order ladder diagrams of the
pion exchange has almost no effect on the form of the nucleon self-energy. 
It appears that the one-pion exchange dominates the whole momentum
dependence in the nucleon self-energy. Although the HF and the DBHF 
self-energies are similar on the scale of some 100 MeV, one should, however,
not conclude that the higher order correlations in the T-matrix are of minor
importance. Since physical observables, like the single particle potential
or the equation-of-state result from the cancelation of the large scalar and
vector fields, they react extremely sensitive on small fluctuations on the 
scale of the self-energy. This is reflected, e.g., by the values of the 
binding energies per particle which is 
(at $\kf=1.34 fm^{-1}$ for full Bonn A) $E/A=-16.9 MeV$ for the DBHF and 
$E/A=+19.3 MeV$ for the pure HF calculation. 
We will now consider the role of the pion in more detail.
\subsection{Conventional pseudo-vector representation}
The discussion of the pion-exchange contribution to the
nucleon self-energy is intimately connected to an ambiguity of the 
T-matrix representation, as it was pointed out in \cite{tjon85a}.
The set of five covariants used to represent the on-shell T-matrix is not 
uniquely defined when one works exclusively in the subspace 
of positive energy states \cite{poschenrieder88}. 
Obviously, various alternative sets of five linearly
independent covariants exist which all can reproduce, like the 
Fermi covariants, the five on-shell helicity matrix elements of the 
T-matrix. For example, the pseudo-vector covariant 
\beq
{\rm PV}={{\not{\tilde{p}}_1^*-\not{\tilde{q}}_1^*}\over {2\tilde{m}^*_F}}
\gamma_5 \otimes 
{{\not{\tilde{p}}_2^*-\not{\tilde{q}}_2^*}\over {2\tilde{m}^*_F}}
\gamma_5 
\label{pseudovector}
\eeq
with $\tilde{q}_i^*$ and $\tilde{p}_i^*$ being the initial and final momenta
of the nucleons, leads to identical on-shell helicity matrix elements as the 
pseudo-scalar covariant ${\rm P}=\gamma_5\otimes \gamma_5$. 
On-shell the nucleon spinors fulfill the quasi-free Dirac equation 
(\ref{dirac2}), with $\tilde{m}^*_F$ fixed at a reference point, 
and therefore it holds 
\beq
\bar{u}(p,\kf){{\not{\tilde{p}}^*-\not{\tilde{q}}^*}\over {{2\tilde{m}^*_F}}}
\gamma_5 u(q,\kf)=\bar{u}(p,\kf)\gamma_5u(q,\kf)
\quad .
\eeq 
Thus, if we replace the pseudo-scalar covariant P 
in the T-matrix representation (\ref{tmatrep1}) by the pseudo-vector 
covariant PV and perform the inversion of 
Eq. (\ref{Tmatinv}), the calculated Lorentz invariant amplitude 
$F_{\rm PV}$ will be identical to the previously calculated amplitude 
$F_{\rm P}$. Hence, the representation of the on-shell T-matrix is 
ambiguous in the detailed form of the covariant operators one can use which is
crucial for the description of the one-pion exchange \cite{fuchs98}.
The $\pi$NN vertex in the OBE potentials, e.g. Bonn, is usually treated by a
pseudo-vector coupling. In \cite{kondratyuk98} in was shown that the $\pi$NN coupling is 
by less than $5 \%$ of pseudo-scalar nature at the on-shell point. There
are several arguments which in addition strongly support a pseudo-vector
vertex. First of all a PV vertex is consistent with soft pion theorems based
on chiral symmetry considerations of $QCD$ \cite{weinberg68}. Secondly, the PV vertex
suppresses the coupling to negative energy states due to the on-shell relation 
\beq
\bar{v}(p,\kf){{\not{\tilde{p}}_1^*-\not{\tilde{q}}_1^*}\over 
{{2\tilde{m}^*_F}}}\gamma_5 u(q,\kf)=0
\quad ,
\eeq 
with $v(p,\kf)$ a negative energy spinor. 
In \cite{tjon85a} it was, e.g., shown that the one-pion exchange contribution to 
the nuclear optical potential tends to increase drastically at low momenta if 
the $\pi$NN vertex is treated as pseudo-scalar. One reason for this behavior is the 
strong coupling to negative energy states which is not apparent in 
non-relativistic approaches. A pseudo-vector vertex is more consistent with 
the approximation scheme of the conventional Brueckner scheme where one neglects 
the negative energy states completely.
It also strongly suppresses the pion contribution in particular 
at low energies which is more in accordance with the empirical knowledge from 
proton-nucleus scattering. One should expect therefore that in the 
DBHF approach a pseudo-vector coupling also drastically reduces 
the influence of the pion on the nucleon self-energy in the 
medium \cite{horowitz87,terhaar87}.

To account for the pseudo-vector nature of the pion exchange in the
T-matrix, in the past the pseudo-scalar covariant P was simply replaced 
by the pseudo-vector covariant PV \cite{terhaar87,sehn97,boersma94,nuppenau89}.
If one uses, however, the 'direct' $ps$ representation of the
T-matrix (\ref{tmatrep1}) and changes to a $pv$ representation, 
the pseudo-vector amplitude $F_{\rm PV}=F_{\rm P}$ will not 
contribute to the self-energy (\ref{self1}) since it does not appear in the
Hartree integral. To suppress the pion contribution 
one needs to find a different $pv$ representation of the T-matrix where 
the pseudo-vector covariant occurs in the Fock or 'exchange' part of the 
self-energy. One possible $pv$ representation is given by the
$ps$ representation (\ref{tmatrep2}) with P and $\tilde{\rm P}$ 
replaced by PV and $\widetilde{\rm PV}$, respectively. The interchanged covariant 
$\widetilde{\rm PV}$ is thereby defined as in \cite{tjon85a} by applying the operator 
$\tilde{S}$, (\ref{excovariants}), to the covariant PV, 
(\ref{pseudovector}), and by interchanging the final momenta 
$\tilde{p}_1^*$ and $\tilde{p}_2^*$ in Eq. (\ref{pseudovector}).
Performing this replacement the pseudo-vector amplitude 
$\bar{F}^X_{\rm PV}=\bar{F}^X_{\rm P}$
contributes to the nucleon self-energy integral with a weight factor of 
\cite{fuchs98}
\beq
Tr[(\not{\tilde{q}}^*+\tilde{m}^*_F)\widetilde{\rm PV}]=-
(\not{\tilde{k}}^*+\tilde{m}^*_F)({{\tilde{k}_\mu^*\tilde{q}^{* \mu}}\over
{2\tilde{m}^*_F}}-{1\over 2})
\quad .
\eeq
For comparison, with the $ps$ representation (\ref{tmatrep2}) the weight factor 
of the pseudo-scalar amplitude $\bar{F}^X_{\rm P}$ in the self-energy integral 
(\ref{self2}) is 
\beq
Tr[(\not{\tilde{q}}^*+\tilde{m}^*_F)\tilde{\rm P}]=-
(\not{\tilde{q}}^*-\tilde{m}^*_F) \quad .
\eeq
Using the $pv$ representation of the T-matrix as discussed above 
the nucleon self-energy becomes \cite{terhaar87,sehn97}
\beqa
 \Sigma_{\alpha\beta}(k,\kf) =  
 \int \frac{d^3{\bf q}}{(2\pi)^3} 
\frac{\Theta(k_F-|{\bf q}|)}{4\tilde{E}^*({\bf q})}
 \left\{ 
({\not {\tilde{k}}}^*_{\alpha\beta}- {\not {\tilde{q}}}^*_{\alpha\beta})
\frac{2 \tilde{q}^*_{\mu} (\tilde{k}^{\ast\mu} 
- \tilde{q}^{*\mu})}{4\tilde{m}^{*2}_F}
                         \bar{F}_{PV}^X \right. \nonumber \\
\left . + \tilde{m}^*_F 1_{\alpha\beta}\left[  
4 \bar{F}_S^D - \bar{F}_S^X -4 \bar{F}_V^X - 12 \bar{F}_T^X + 4 \bar{F}_A^X  
 - \frac{(\tilde{k}^{*\mu} - \tilde{q}^{*\mu})^2}{4\tilde{m}^{*2}_F}
\bar{F}_{PV}^X\right]    \right. \nonumber \\
\left.  + {\not{\tilde{q}}}^{*}_{\alpha\beta} 
\left[ 4 \bar{F}_V^D - \bar{F}_S^X +2 \bar{F}_V^X+ 2 \bar{F}_A^X  
 - \frac{(\tilde{k}^{*\mu} - \tilde{q}^{*\mu})^2}{4\tilde{m}^{*2}_F}
\bar{F}_{PV}^X\right] \right\} .
\label{self3}
\eeqa
However, this self-energy appears to be unphysical since 
the amplitudes $\bar{F}^{D,X}_i$ used in the integral are calculated from 
non-antisymmetrized helicity matrix elements of the T-matrix. 
It is by no means clear if the unphysical contributions do cancel as 
they do in the case when we use the $ps$ representation of the T-matrix, 
i.e. see integrals (\ref{self2}) and (\ref{self1}). 
\begin{figure}[h]
\vspace{75mm}
\includegraphics{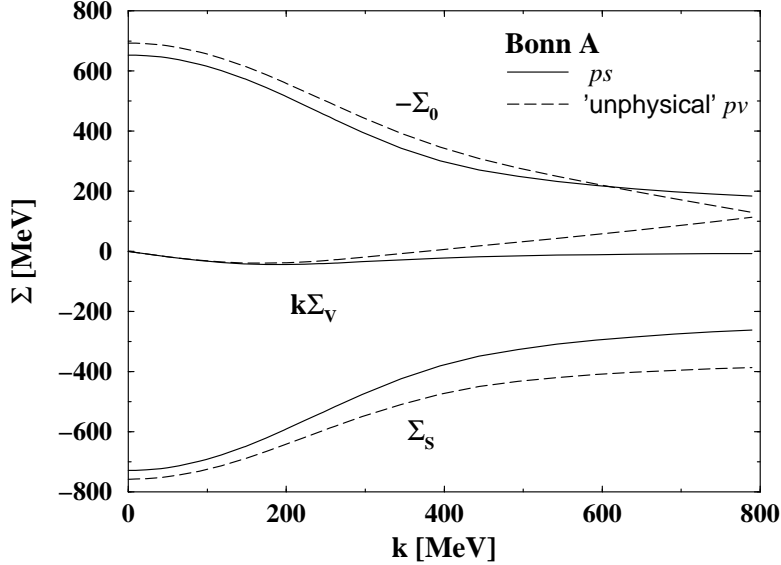}
\caption{\label{fig3} Momentum dependence of the DBHF nucleon self-energy 
components in nuclear matter at $\kf=1.34 fm^{-1}$ using the Bonn A 
potential. For the T-matrix the $ps$ representation (Eq. (\ref{tmatrep1}), solid) 
and the 'unphysical' $pv$ representation 
(as used in (\ref{self3}), dashed) are applied.}
\end{figure}
That the current $pv$ representation of the T-matrix is indeed unphysical can 
be seen in Fig. \ref{fig3} where the nucleon self-energy components, evaluated 
from the integral (\ref{self3}), are shown. 
The self-energy components continue to be strongly momentum
dependent. Furthermore, the vector component $\Sigma_V$ shows an 
unphysical asymptotic behavior. Instead of dropping to zero 
its value increases with increasing momentum of the nucleon. 
This demonstrates that the $pv$ representation of the T-matrix 
with non-antisymmetrized Lorentz invariant amplitudes 
$\bar{F}_i^{{\rm I}}$, as used in Eq. (\ref{self3}), is not useful.

To circumvent the problem of unphysical contribution to the nucleon 
self-energy one should start from a different $ps$ representation of the
T-matrix where one uses anti-symmetrized amplitudes 
$F_i^{\rm I}(\pabs,\theta,x)$ and $F_i^{\rm I}(\pabs,\pi-\theta,x)$ 
instead of unphysical amplitudes 
$\bar{F}_i^{\rm I}(\pabs,\theta,x)$ and 
$\bar{F}_i^{\rm I}(\pabs,\pi-\theta,x)$, respectively. 
One possible $ps$ representation of the T-matrix is, for example, 
given by \cite{sehn97}
\beq
T^{{\rm I}}(\pabs,\theta,x)=T^{{\rm I},D}(\pabs,\theta,x)
-T^{{\rm I},X}(\pabs,\theta,x)
\quad .
\eeq
where the 'direct' and 'exchange' parts of the T-matrix are defined as    
\beqa
T^{{\rm I},D}(\pabs,\theta,x)&=& {1\over 2} \left[ 
F_{\rm S}^{\rm I}(\pabs,\theta,x){\rm S}+
F_{\rm V}^{\rm I}(\pabs,\theta,x){\rm V}+
F_{\rm T}^{\rm I}(\pabs,\theta,x){\rm T} \right.
\nonumber \\
&+&\left. F_{\rm A}^{\rm I}(\pabs,\theta,x){\rm A}+
F_{\rm P}^{\rm I}(\pabs,\theta,x){\rm P} \right] 
\label{tdir} \quad ,
\eeqa
and 
\beqa
&&T^{{\rm I},X}(\pabs,\theta,x)= (-1)^{\rm I+1}{1\over 2}\left[ 
F_{\rm S}^{\rm I}(\pabs,\pi-\theta,x)\tilde{{\rm S}}+
F_{\rm V}^{\rm I}(\pabs,\pi-\theta,x)\tilde{{\rm V}}
\right. \nonumber \\
&&\left. \,\,\,\,\,\,\,\,
+F_{\rm T}^{\rm I}(\pabs,\pi-\theta,x)\tilde{{\rm T}} 
+F_{\rm A}^{\rm I}(\pabs,\pi-\theta,x)\tilde{{\rm A}}
+F_{\rm P}^{\rm I}(\pabs,\pi-\theta,x)\tilde{{\rm P}} \right]
\label{tex}.
\eeqa
Due to the anti-symmetry relation (\ref{antisymeq1}) for the 
Lorentz invariant amplitudes $F_i^{\rm I}(\pabs,\theta,x)$ we have
the identity
\beq
T^{{\rm I},X}(\pabs,\theta,x)=-T^{{\rm I},D}(\pabs,\theta,x)
\label{tmatidentity}
\eeq
and with the normalization factors ${1\over 2}$ in (\ref{tdir}) and 
(\ref{tex}) this leads to identical results for the self-energies 
as in the former cases, i.e. Eqs. (\ref{self1}) and (\ref{self2}).

\begin{figure}[h]
\vspace{75mm}
\includegraphics{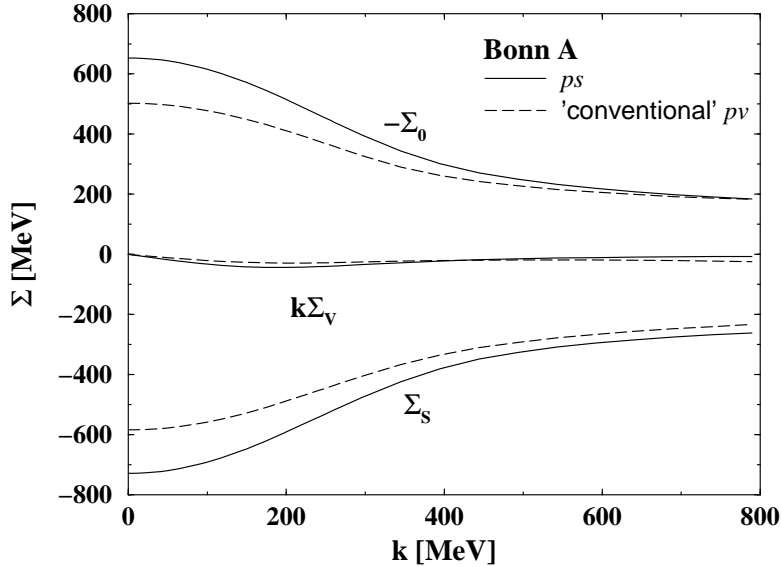}
\caption{\label{fig4} Momentum dependence of the 
DBHF nucleon self-energy components in nuclear matter at 
$\kf=1.34 fm^{-1}$ using the Bonn A potential. For the  T-matrix
the $ps$ representation (Eq. (\ref{tmatrep1}), solid) and 
the 'conventional' $pv$ representation 
(see text after Eq. (\ref{tmatidentity}), dashed) are applied.}
\end{figure}

If we replace now in (\ref{tdir}) and (\ref{tex}) the covariants 
${\rm P},\tilde{\rm P}$ by ${\rm PV},\widetilde{\rm PV}$, respectively, 
we arrive at the 'conventional' $pv$ representation as applied in Refs. 
\cite{terhaar87,sehn97,terhaar87b}.
The nucleon self-energy components calculated with 
this 'conventional' $pv$ representation of the T-matrix are shown 
in Fig. \ref{fig4}.
The behavior of the nucleon self-energy components 
is similar to the case with the $ps$ representation for the T-matrix, i.e. the 
momentum dependence is still pronounced, although the absolute magnitude
of the fields is reduced. 
As discussed in Ref. \cite{fuchs98} 
this is due to the fact that the pion contribution is 
yet not correctly treated as pseudo-vector.
\subsection{Complete pseudo-vector representation}
To suppress the undesirable pseudo-scalar contribution of the pion to the 
nucleon self-energy we have to determine a different $pv$ representation 
of the T-matrix. As starting point we should consider the Hartree-Fock level 
in more detail since the correct $pv$ representation of the T-matrix should
reproduce the HF nucleon self-energy when the pseudo-vector pion exchange 
potential is employed as bare nucleon-nucleon interaction.
The 'conventional' $pv$ representation discussed in the last section does 
not fulfill this minimal requirement. The reason is simply that due to the 
Fierz transformation (\ref{fierz}) all Fermi covariants still 
contain pseudo-scalar contribution, i.e. the covariants 
(S,V,T,A,P) can be re-expressed in terms of 
$(\tilde{\rm S},\tilde{\rm V},\tilde{\rm T},\tilde{\rm A},\tilde{\rm P})$.  
Hence, in the 'conventional' $pv$ representation not all possible 
pseudo-scalar contributions are replaced by a pseudo-vector contribution 
when one simply replaces $\tilde{\rm P}$ with $\widetilde{\rm PV}$ 
in Eq. (\ref{tex}).  
To obtain a 'complete' $pv$ representation the identities 
\beqa 
\hspace{3cm}
\frac{1}{2}({\rm T}+\tilde{\rm T})&=&{\rm S}+\tilde{\rm S}+{\rm P}
+\tilde{\rm P}\\
{\rm V}+\tilde{\rm V}&=&{\rm S}+\tilde{\rm S}-{\rm P}-\tilde{\rm P}
\label{identi}
\eeqa 
are actually very helpful. Since Tjon and Wallace already addressed this 
point in Ref. \cite{tjon85a} we will follow now partially their notation.
To obtain the 'complete' $pv$ representation these authors started 
from a 'symmetrized' $ps$ representation of the form
\beqa
&&T^{\rm I}(\pabs,\theta,x)=
f_1^{\rm I}(\pabs,\theta,x)({\rm S}-\tilde{\rm S})
+f_2^{\rm I}(\pabs,\theta,x){1\over 2}({\rm T}+\tilde{\rm T})
\nonumber \\
&&\,\,\,\,\,\,
-f_3^{\rm I}(\pabs,\theta,x)({\rm A}-\tilde{\rm A})
+f_4^{\rm I}(\pabs,\theta,x)({\rm V}+\tilde{\rm V}) 
+f_5^{\rm I}(\pabs,\theta,x)({\rm P}-\tilde{\rm P})
\label{tmatrep4}
\eeqa
for the T-matrix. Again, this $ps$ representation is equivalent to the 
$ps$ representation (\ref{tmatrep1}). Due to the  
Fierz transformation (\ref{fierz}) the five amplitudes 
$f_i^{\rm I}$ are related to the amplitudes $F_i^{\rm I}$ by a linear 
transformation 
\beqa
\hspace{2cm}
\left( 
\begin{array}{c} 
f_1^{\rm I} \\ 
f_2^{\rm I} \\ 
f_3^{\rm I} \\ 
f_4^{\rm I} \\ 
f_5^{\rm I} 
\end{array} 
\right)
= {1\over 4}
\left( 
\small{ 
\begin{array}{ccccc} 
 2 & -4 & -12 & 0  &  0  \\
 1 &  0 &  4  & 1  &  0  \\
 0 & -2 &  0  & 0  &  -2 \\
 1 &  2 &  0  & -1 &  -2 \\
 0 &  4 & -12 & 2  &  0 
\end{array}} 
\right)
\left( 
\begin{array}{c} 
F_{\rm S}^{\rm I} \\ 
F_{\rm V}^{\rm I} \\ 
F_{\rm T}^{\rm I} \\ 
F_{\rm P}^{\rm I} \\ 
F_{\rm A}^{\rm I} 
\end{array} 
\right) .
\label{transform1}
\eeqa
The anti-symmetry relation (\ref{antisymeq1}) converts to the much  
simpler phase relation \cite{tjon85a}
\beq
f^{\rm I}_i (\pabs,\pi-\theta,x) = (-)^{\rm I+ i}f^{\rm I}_i (\pabs,\theta,x)
\quad ,
\eeq
where i runs from 1 to 5. 
Applying the operator identities (\ref{identi}) the 'symmetrized' 
$ps$ representation (\ref{tmatrep4}) can be rewritten as
\beqa
\hspace{1cm}
T^{\rm I}(\pabs,\theta,x)&=& g_{\rm S}^{\rm I}(\pabs,\theta,x){\rm S}
-g_{\tilde{\rm S}}^{\rm I}(\pabs,\theta,x)\tilde{\rm S}
+g_{\rm A}^{\rm I}(\pabs,\theta,x)({\rm A}-\tilde{\rm A})\nonumber \\ 
&+& g_{\rm P}^{\rm I}(\pabs,\theta,x){\rm P}
-g_{\tilde{\rm P}}^{\rm I}(\pabs,\theta,x)\tilde{\rm P}
\quad ,
\label{tmatrep5}
\eeqa
where the new amplitudes $g_i^{\rm I}$ are defined as 
\beqa
\hspace{2cm}
\left( 
\begin{array}{c} 
g_{\rm S}^{\rm I} \\ 
g_{\tilde{\rm S}}^{\rm I} \\ 
g_{\rm A}^{\rm I} \\ 
g_{\rm P}^{\rm I} \\ 
g_{\tilde{\rm P}}^{\rm I} 
\end{array} 
\right)
= 
\left( 
\small{ 
\begin{array}{ccccc} 
 1 &  1 &  0  & 1  &  0  \\
 1 & -1 &  0  & -1 &  0  \\
 0 &  0 &  -1 & 0  &  0 \\
 0 &  1 &  0  & -1 &  1 \\
 0 & -1 &  0  & 1  &  1 
\end{array}} 
\right)
\left( 
\begin{array}{c} 
f_1^{\rm I} \\ 
f_2^{\rm I} \\ 
f_3^{\rm I} \\ 
f_4^{\rm I} \\ 
f_5^{\rm I} 
\end{array} 
\right) .
\label{transform2}
\eeqa
The relation between the amplitudes $g_i^{\rm I}$ and $F_i^{\rm I}$ 
is given by the matrix product of the linear transformations 
(\ref{transform1}) and (\ref{transform2}), i.e. 
\beqa
\hspace{2cm}
\left( 
\begin{array}{c} 
g_{\rm S}^{\rm I} \\ 
g_{\tilde{\rm S}}^{\rm I} \\ 
g_{\rm A}^{\rm I} \\ 
g_{\rm P}^{\rm I} \\ 
g_{\tilde{\rm P}}^{\rm I} 
\end{array} 
\right)
= {1\over 4}
\left( 
\small{ 
\begin{array}{ccccc} 
 4 & -2 & -8  & 0  & -2  \\
 0 & -6 & -16 & 0  &  2  \\
 0 & -2 &  0  & 0  & -2 \\
 0 &  2 & -8  & 4  &  2 \\
 0 &  6 & -16 & 0  & -2 
\end{array}} 
\right)
\left( 
\begin{array}{c} 
F_{\rm S}^{\rm I} \\ 
F_{\rm V}^{\rm I} \\ 
F_{\rm T}^{\rm I} \\ 
F_{\rm P}^{\rm I} \\ 
F_{\rm A}^{\rm I} 
\end{array} 
\right) .
\label{transform3}
\eeqa
Due to the linear relations between the amplitudes
$F_i^{\rm I}$, $f_i^{\rm I}$ and $g_i^{\rm I}$, all three 
$ps$ representations (\ref{tmatrep1}), (\ref{tmatrep4}) and
(\ref{tmatrep5}) of the T-matrix lead to identical results 
for the nucleon self-energy. Using the representation (\ref{tmatrep5})
the self-energy integral reads simply 
\beqa
\hspace{1cm}
\Sigma_{\alpha\beta}(k,\kf)&=& \int {{d^3{\bf q}}\over {(2\pi)^3}}
{{\theta(\kf-|{\bf q}|)}\over {4\tilde{E}^*({\bf q})}}
\left[\tilde{m}^*_F 1_{\alpha\beta} 
\left(
4 g_{\rm S}-g_{\tilde{\rm S}}
+4 g_{\rm A}-g_{\tilde{\rm P}} \right)
\right. \nonumber \\
&&\left.
+ \not{\tilde q}^*_{\alpha\beta}
\left(
-g_{\tilde{\rm S}}+2g_{\rm A}
+g_{\tilde{\rm P}}\right)\right]
\quad ,
\label{self4}
\eeqa
where again $g_i$ denote isospin averaged amplitudes, 
evaluated at relative momenta $|\bf p|$ and 
scattering angle $\theta=0$ in the c.m. frame.

\begin{figure}[h]
\vspace{145mm}
\includegraphics{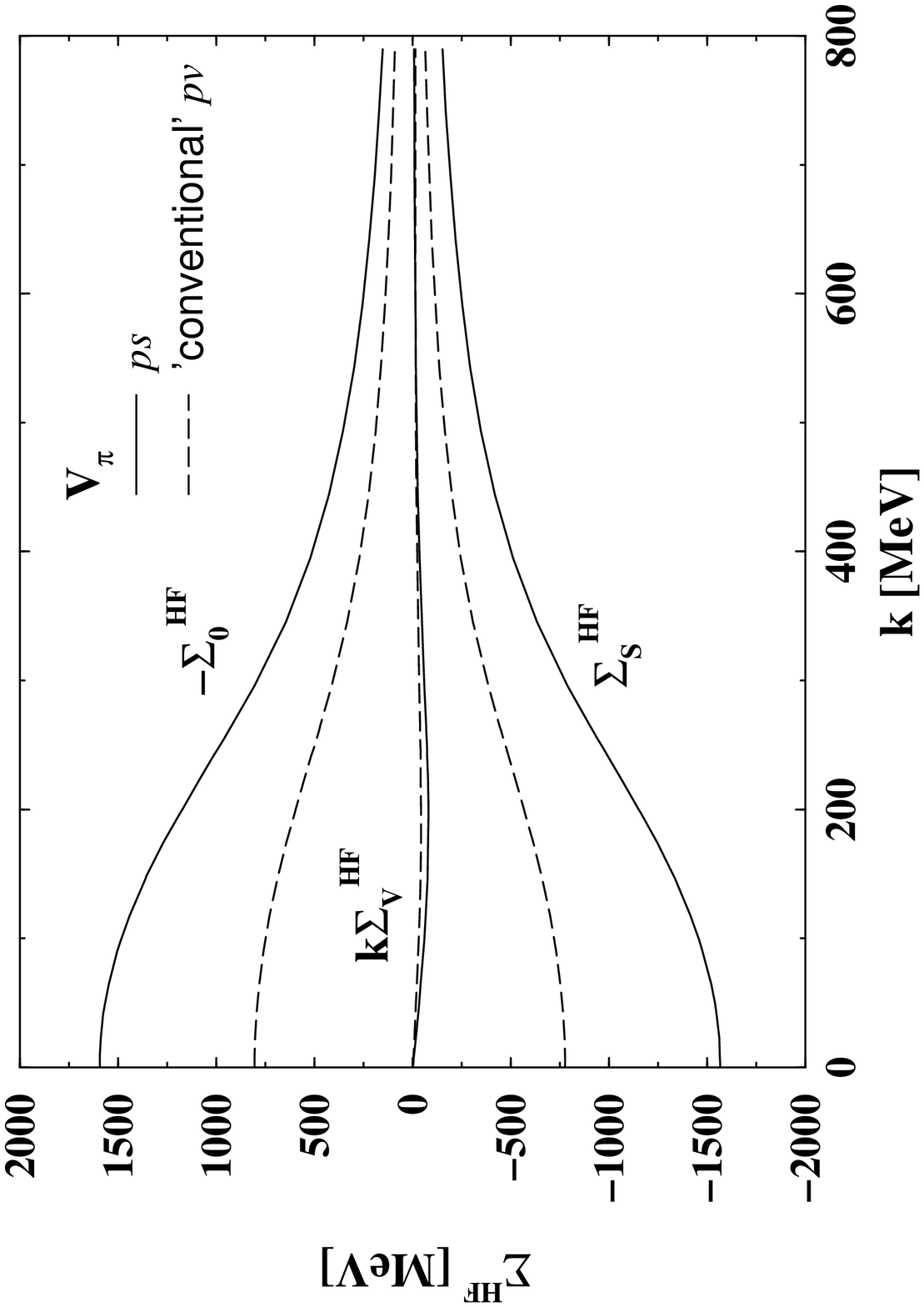}
\includegraphics{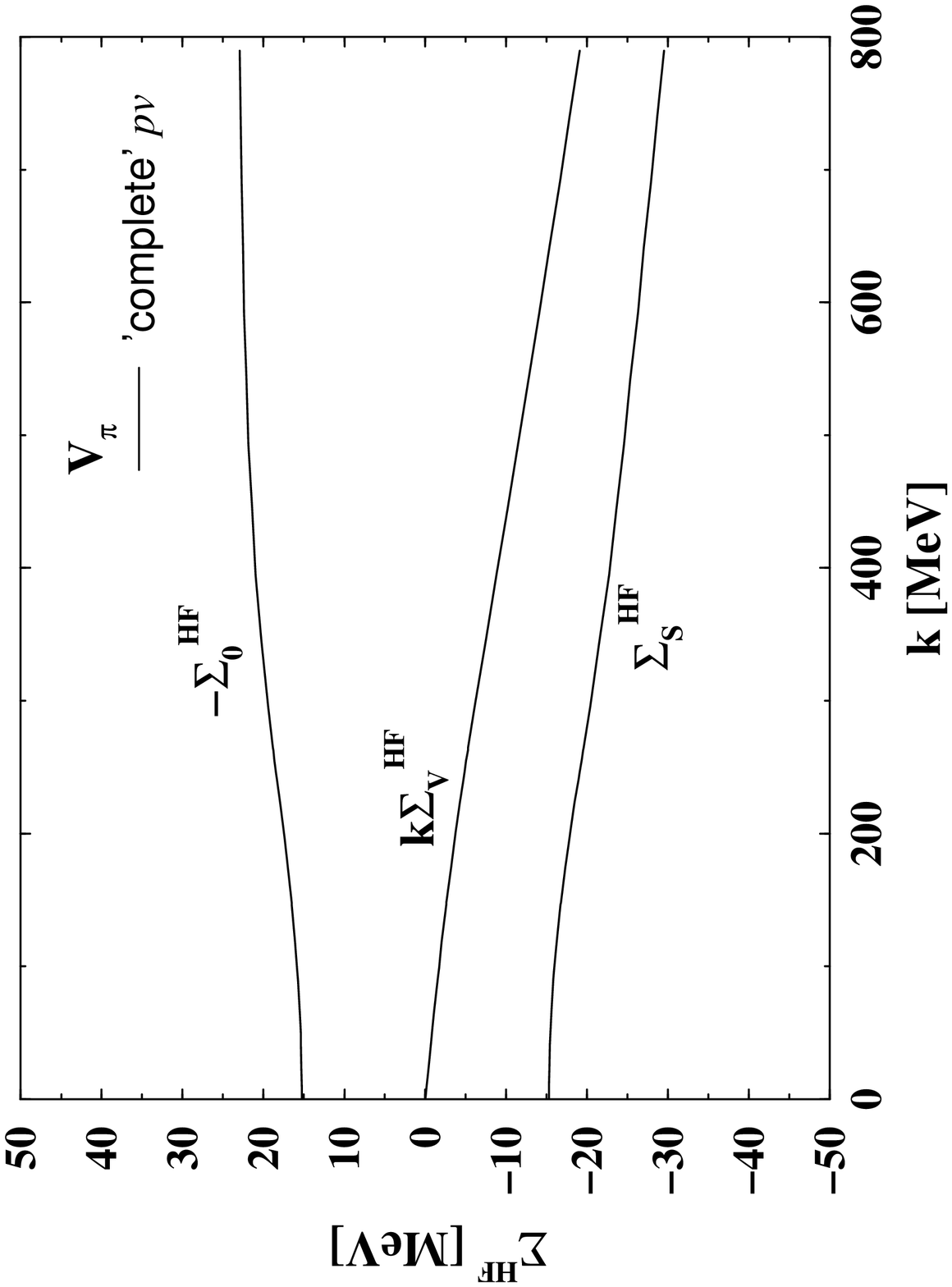}
\caption{\label{fig5} Momentum dependence of the non-selfconsistent 
Hartree-Fock nucleon self-energy components in nuclear matter 
at $\kf=1.34 fm^{-1}$. As bare nucleon-nucleon interaction the single-pion
exchange potential is used. In the upper panel the 
$ps$ representation (Eq. (\ref{tmatrep1}, solid) and the 
'conventional' $pv$ representation (see text after 
Eq. (\ref{tmatidentity}), dashed) of the interaction are used. The lower panel
shows the result using the 'complete' $pv$ representation 
(Eq. (\ref{tmatrep6}), solid) of the interaction.}
\end{figure}

If we replace in (\ref{tmatrep5}) the covariants 
${\rm P}$, $\tilde{\rm P}$ by the pseudo-vector covariants ${\rm PV}$, 
$\widetilde{\rm PV}$, respectively, we arrive at the 
'complete' $pv$ representation \cite{tjon85a}
\beqa
\hspace{1cm}
T^{\rm I}(\pabs,\theta,x)&=& g_{\rm S}^{\rm I}(\pabs,\theta,x){\rm S}
-g_{\tilde{\rm S}}^{\rm I}(\pabs,\theta,x)\tilde{\rm S}
+g_{\rm A}^{\rm I}(\pabs,\theta,x)({\rm A}-\tilde{\rm A})\nonumber \\
&+& g_{\rm PV}^{\rm I}(\pabs,\theta,x){\rm PV}
-g_{\widetilde{\rm PV}}^{\rm I}(\pabs,\theta,x)\widetilde{\rm PV}
\quad ,
\label{tmatrep6}
\eeqa
with $g_{\rm PV}^{\rm I}(\theta)$ and $g_{\widetilde{\rm PV}}^{\rm I}(\theta)$ 
being identical to $g_{\rm P}^{\rm I}(\theta)$ and 
$g_{\tilde{\rm P}}^{\rm I}(\theta)$, respectively. 
As shown in Ref. \cite{fuchs98}, this representation is able to reproduce 
the Hartree-Fock results for the nucleon self-energy 
when we use the pseudo-vector pion exchange potential as bare nucleon-nucleon 
interaction. The self-energy integral using this 'complete' 
$pv$ representation of the T-matrix is given by 
\beqa
 \Sigma_{\alpha\beta}(k,\kf) =  
 \int \frac{d^3{\bf q}}{(2\pi)^3} 
\frac{\Theta(k_F-|{\bf q}|)}{4\tilde{E}^*({\bf q})}
 \left\{ 
({\not {\tilde{k}}}^*_{\alpha\beta}- {\not {\tilde{q}}}^*_{\alpha\beta})
\frac{2 \tilde{q}^*_{\mu} (\tilde{k}^{\ast\mu} 
- \tilde{q}^{*\mu})}{4\tilde{m}^{*2}_F}
g_{\widetilde{\rm PV}}\right. \nonumber \\
\left . + \tilde{m}^*_F 1_{\alpha\beta}\left[  
4 g_{\rm S} - g_{\tilde{\rm S}}  + 4 g_{\rm A}  
 - \frac{(\tilde{k}^{*\mu} - \tilde{q}^{*\mu})^2}{4\tilde{m}^{*2}_F}
g_{\widetilde{\rm PV}} \right]    \right. \nonumber \\
\left.  + {\not{\tilde{q}}}^{*}_{\alpha\beta} 
\left[ - g_{\tilde{\rm S}}+2 g_{\rm A}  
 - \frac{(\tilde{k}^{*\mu} - \tilde{q}^{*\mu})^2}{4\tilde{m}^{*2}_F}
g_{\widetilde{\rm PV}} \right] \right\} 
\quad .
\label{self5}
\eeqa
In Fig. \ref{fig5} we show the non-selfconsistent Hartree-Fock nucleon 
self-energy components in nuclear matter using only the pion exchange potential
as bare interaction between the nucleons. 
In the upper panel the $ps$ representation as well as 
the 'conventional' $pv$ representation are used 
while the lower panel shows the results 
with the 'complete' $pv$ representation of the interaction. 
With the $ps$ representation and the 'complete'
$pv$ representation one can reproduce the analytical results for the 
Hartree-Fock nucleon self-energies if we use either the pseudo-scalar or the
pseudo-vector $\pi$NN vertex for the pion exchange potential, respectively.
Hence, the 'complete' $pv$ representation is the correct pseudo-vector 
representation for the interaction.
On the other hand the 'conventional' $pv$ representation leads to 
wrong results for the HF nucleon self-energy. In this representation the 
pion is not completely treated as pseudo-vector. 
Fig. \ref{fig5} also shows the influence of the single-pion
exchange potential to the nucleon self-energy. Only when we use the 
'complete' $pv$ representation the contribution of the pion to the 
nucleon self-energy is weak. In all other  
cases, pseudo-scalar or 'conventional' pseudo-vector, the influence of the pion 
is extremely strong, i.e. the contribution to the self-energy is at least  
one or two orders of magnitude larger than in the complete pseudo-vector case.

\begin{figure}[h]
\vspace{75mm}
\includegraphics{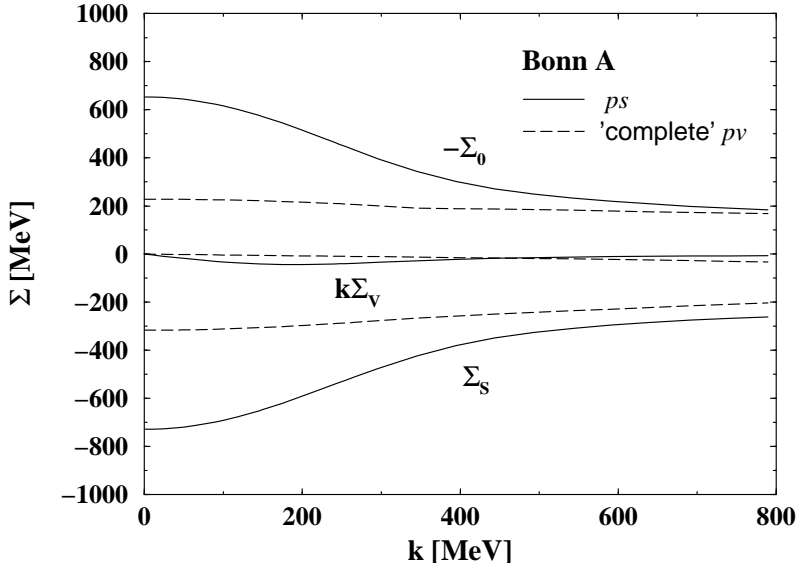}
\caption{\label{fig6} Momentum dependence of the DBHF nucleon self-energy
components in nuclear matter at $\kf=1.34 fm^{-1}$ 
using Bonn A as bare nucleon-nucleon interaction. For the 
T-matrix the $ps$ representation (Eq. (\ref{tmatrep1}), solid) and the 
'complete' $pv$ representation (Eq. (\ref{tmatrep6}), dashed) are applied.}
\end{figure}

In Fig. \ref{fig6} we present the full self-consistent DBHF calculation 
with the 'compete' $pv$ representation of the T-matrix \cite{fuchs98}. 
The DBHF nucleon self-energy components are indeed weakly momentum dependent.
The single-pion exchange contribution to the interaction, 
which was previously dominating at low nucleon momenta, 
is now strongly suppressed.  
Consequently the result within the 'complete' pv representation using the
Bonn A potential resembles the result within the $\sigma$-$\omega$ model 
potential, see Fig. \ref{fig1} where the $ps$ representation was used. 
To suppress the pion contribution to the in-medium T-matrix 
a correct pseudo-vector like covariant representation is essential for the 
calculation of the nucleon self-energy in nuclear matter. 
As it is necessary for the whole calculation scheme, the weak momentum 
dependence of the nucleon self-energy is also in accordance with the 
'reference spectrum approximation' used in the calculation. 
The current DBHF approach therefore appears to be self-consistent.
\subsection{Covariant representations of the subtracted T-matrix}
\begin{figure}[h]
\vspace{75mm}
\includegraphics{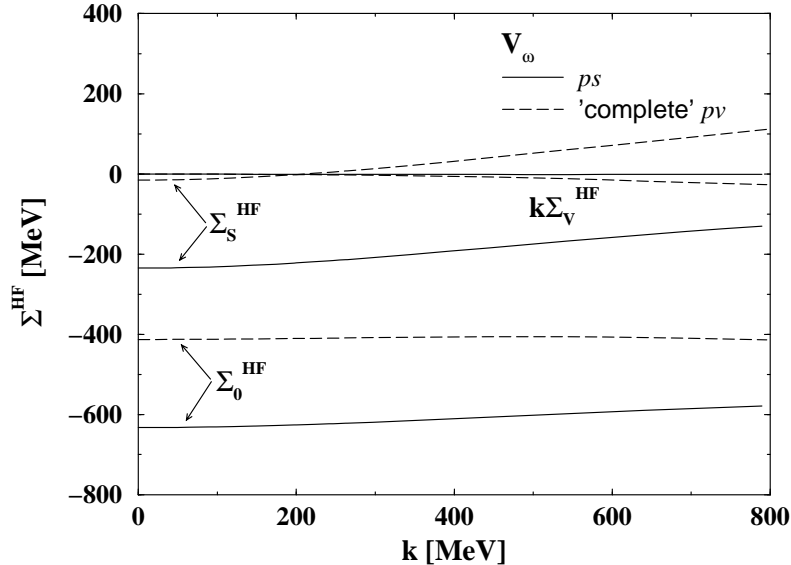}
\caption{\label{fig7} Momentum dependence of the 
non-selfconsistent Hartree-Fock nucleon self-energy components in nuclear
matter at $\kf=1.34 fm^{-1}$ using as bare interaction the omega exchange 
potential. For the potential the $ps$ representation 
(Eq. (\ref{tmatrep1}), solid) and the 'complete' $pv$ representation 
(Eq. (\ref{tmatrep6}), dashed) are used.}
\end{figure}

The 'complete' $pv$ representation successfully 
reproduces the HF nucleon self-energy in the case of the pion exchange.
Hence, this representation is at the moment the 'best' representation 
of the on-shell T-matrix which is accordance with the pseudo-vector nature of 
the pion exchange potential.
However, as already pointed out in \cite{fuchs98}, the 'complete' 
$pv$ representation fails to reproduce the HF nucleon self-energy if other 
meson exchange potentials are applied as bare interaction. 
This is demonstrated in Fig. \ref{fig7} where we consider as an example 
the single-omega exchange. 
While the $ps$ representation of the interaction correctly 
reproduces the analytical HF nucleon self-energy, the 
'complete' $pv$ representation fails in this respect.
In particular the scalar and vector self-energies are shifted by about
200 MeV. 
The failure of the 'complete' $pv$ representation 
is understandable since the vector covariant V in the 
$ps$ representation is partially replaced by a pseudo-vector covariant 
when we use the identity (\ref{identi}) with P and $\tilde{\rm P}$ 
replaced by PV and $\widetilde{\rm PV}$, respectively. 
Since the 'complete' $pv$ representation is not the 
correct covariant representation of the bare interaction,  
we therefore can not expect that 
it is correct on the level of the full in-medium interaction. 

However, in the last section we have seen that the influence of the 
pion is dominantly given by the single-pion exchange. 
Hence, it should be reasonable to treat the bare interaction and 
the higher order ladder graphs of the meson exchange potential separately. 
Since the single-meson exchange potential is actually known analytically 
we can represent it covariantly by a mixed representation of the form 
\beq
V=V_{\pi,\eta}^{PV}+V_{\sigma,\omega,\rho,\delta}^{P}
\quad .
\label{mixed1}
\eeq
Here the $\pi$- and $\eta$-meson contributions  
are treated as pseudo-vector (\ref{tmatrep5})
while for the ($\sigma,\omega,\rho,\delta$)-meson contributions 
of the Bonn potential the $ps$ representation (\ref{tmatrep4})
is applied. The higher order ladder diagrams of the T-matrix 
\beq
T_{Sub}=T-V=i\int VQGGT = \sum\limits_{n=1}^\infty \int V(iQGGV)^n
\quad ,
\eeq
in the following called the subtracted T-matrix, can not be represented 
correctly in a mixed form since we can not disentangle the different
meson contributions to this part of the full in-medium interaction.
The representation of the subtracted T-matrix remains therefore ambiguous.
However, if the pion exchange dominantly contributes to 
the Hartree-Fock level a $ps$ representation of the subtracted T-matrix 
should be more appropriate because then the higher order contributions of 
other meson exchange potentials are not treated incorrectly as pseudo-vector. 
Thus the most favorable representation of the T-matrix is given by the 
$ps$ representation
\beq
T^{\rm P}=T^{\rm P}_{Sub}+V_{\pi,\eta}^{PV}+V_{\sigma,\omega,\rho,\delta}^{P}
\quad .
\label{tmatps}
\eeq
Here the $ps$ representation for $T^{\rm P}_{Sub}$ is determined via the 
matrix elements
\beqa
<{\bf p}\lambda_1^{'}\lambda_2^{'}|T^{\rm I}_{Sub}(x)|
{\bf q}\lambda_1\lambda_2>:=
<{\bf p}\lambda_1^{'}\lambda_2^{'}|T^{\rm I}(x)-V^{\rm I}(x)|
{\bf q}\lambda_1\lambda_2> 
\quad ,
\label{tmatsub}
\eeqa
with subsequently applying the projection scheme as in Eq. (\ref{tmatinv}).
An alternative representation of the T-matrix is given
by the $pv$ representation
\beq
T^{\rm PV}=T^{\rm PV}_{Sub}+V_{\pi,\eta}^{PV}+V_{\sigma,\omega,\rho,\delta}^{P}
\quad ,
\label{tmatpv}
\eeq
where the subtracted T-matrix is represented by the 'complete' 
$pv$ representation (\ref{tmatrep6}). This representation is similar
to the 'complete' $pv$ representation of the full T-matrix, however, with the
advantage that now the pseudo-scalar contributions in the bare 
nucleon-nucleon interaction, e.g. the single-omega exchange potential, 
are represented correctly. In the next section we will use both 
representations, (\ref{tmatps}) and (\ref{tmatpv}), to study the properties 
of nuclear matter in the DBHF approach. 
In this way we can determine the influence of the higher order 
ladder graphs to the in-medium interaction in a more quantitative way. 
Furthermore, these two representations set the range of the remaining 
ambiguity concerning the representation of the T-matrix, i.e. after
separating the leading order contributions.
\section{Results for nuclear matter}
In this section we will present results for the properties of nuclear matter 
using the new approach to the DBHF problem outlined in the previous section. 
As bare nucleon-nucleon potential we employ the one-boson exchange potentials   
Bonn A, B and C \cite{machleidt86}. These potentials are based on the exchange of six 
non-strange bosons ($\pi,\eta,\rho,\omega,\delta,\sigma$) with masses 
below 1 GeV. For the pion and the eta meson the derivative pseudo-vector 
coupling is applied. The three parameterizations A,B and C of the Bonn potential
differ essentially in the $\pi$NN form factor and, as a consequence, 
in the strength of the nuclear tensor force. 
\subsection{The nucleon self-energy}
\subsubsection{Momentum dependence}
\begin{figure}[h]
\vspace{75mm}
\includegraphics{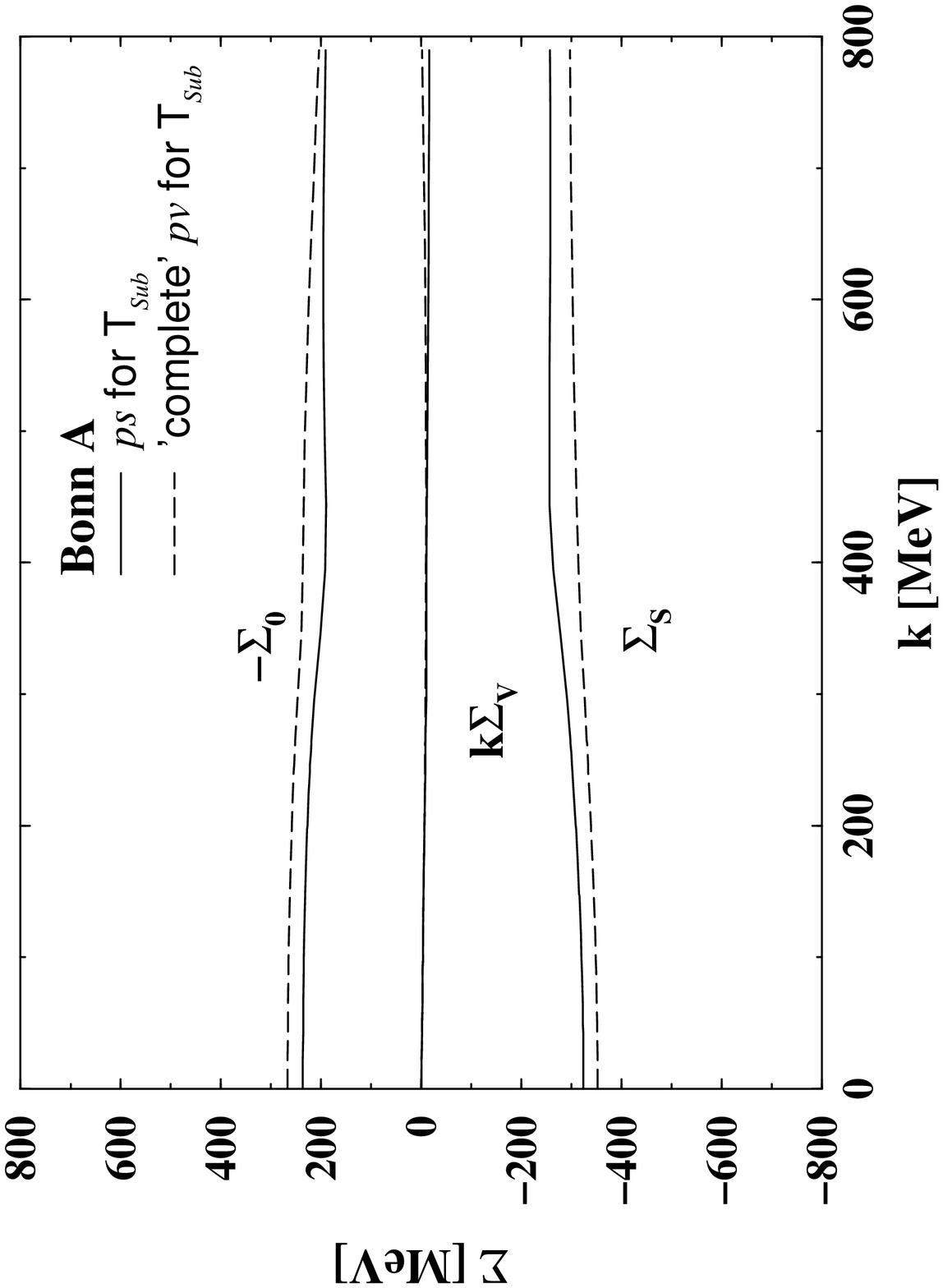}
\caption{\label{fig8} Momentum dependence of the nucleon self-energy
components in nuclear matter at $\kf=1.34 fm^{-1}$ using Bonn A as 
bare nucleon-nucleon interaction. For the 
T-matrix the subtraction scheme with the $ps$ representation 
(Eq. (\ref{tmatps}), solid) 
and the $pv$ representation (Eq. (\ref{tmatpv}), dashed) are applied.}
\end{figure}

The momentum dependence of the nucleon self-energy 
at saturation density in isospin saturated nuclear matter is shown in 
Fig. \ref{fig8}. 
For both T-matrix representations (\ref{tmatps})
and (\ref{tmatpv}), the self-energy components are rather weakly dependent 
on the nucleon momentum. In addition,
they are also almost identical, i.e. the difference between a
pseudo-scalar or pseudo-vector representation of the higher order ladder diagrams is
rather small. This demonstrates again that the pion indeed 
contributes mostly to the Hartree-Fock level. 
Hence the ambiguity of the T-matrix representation has only minor
influence on the final result for the nucleon self-energy in the medium

\begin{figure}[h]
\vspace{75mm}
\includegraphics{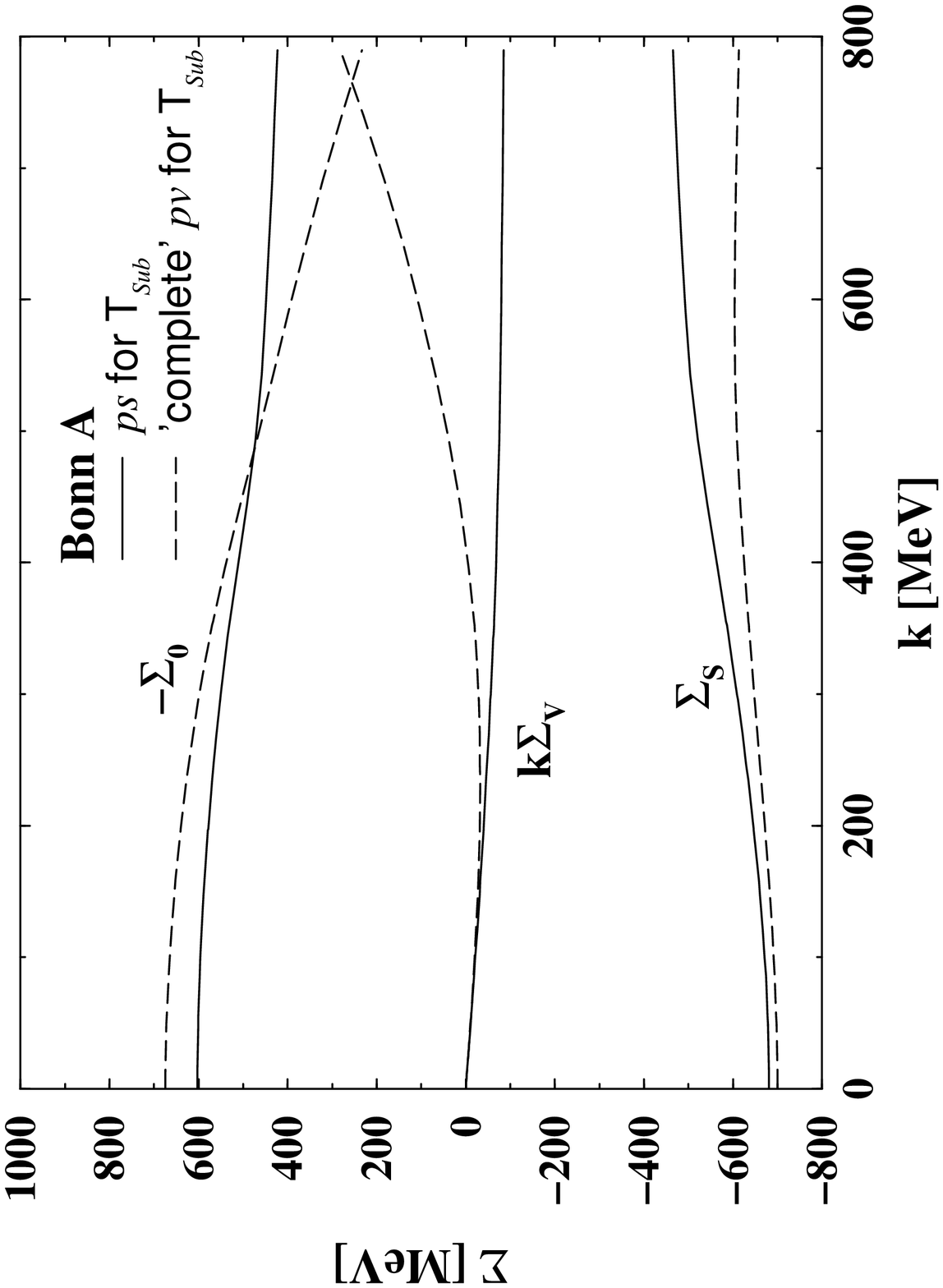}
\caption{\label{fig9} Momentum dependence of the nucleon self-energy
components in nuclear matter at $\kf=1.8 fm^{-1}$ using Bonn A as 
bare nucleon-nucleon interaction. For the T-matrix the subtraction scheme 
with the $ps$ representation (Eq. (\ref{tmatps}), solid) and the 
$pv$ representation (Eq. (\ref{tmatpv}), dashed) are applied.}
\end{figure}

However, at larger densities of the nuclear medium the situation changes. 
In Fig. \ref{fig9} the momentum dependence of the 
self-energy components in nuclear matter at a Fermi momentum of 
$\kf=1.8 fm^{-1}$ is shown.
Now the two representations of the T-matrix lead to significantly 
different results. 
In addition, the momentum dependence of the nucleon self-energy components is 
increasing for both representations.
A strong momentum dependence of the self-energies at higher densities was also 
observed in Ref. \cite{terhaar87}.  The question how to represent the 
T-matrix is now much more severe than at lower densities. 
However, the 'complete' $pv$ representation of the 
subtracted T-matrix gives a rather 
unphysical asymptotic behavior of the nucleon self-energy. 
The vector component $\Sigma_V$ not only changes 
sign but also increases drastically at large nucleon momenta. 
We believe that the misrepresentation of the higher order ladder diagrams
of the heavy meson exchange potentials which should be treated as 
pseudo-scalar and not as pseudo-vector is responsible for this behavior.
Since the $ps$ representation of the ladder kernel,
i.e. the subtracted T-matrix, still yields reasonable
results at higher densities this representation should be preferable. 
Since the momentum dependence increases with increasing density, it should be 
included in the future in the self-consistency scheme when predictions 
at high densities are made \cite{lee97}.

\begin{figure}[h]
\vspace{145mm}
\includegraphics{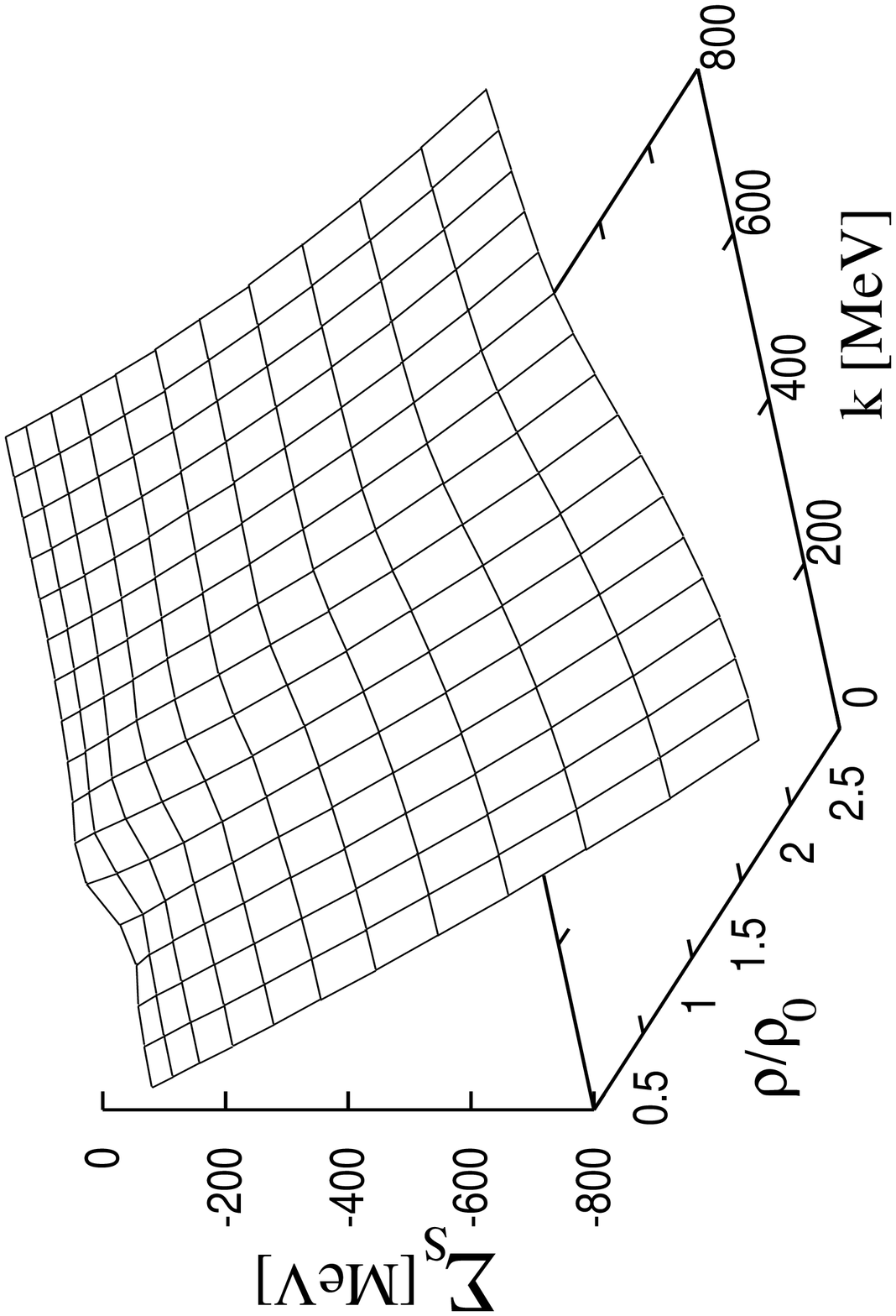}
\includegraphics{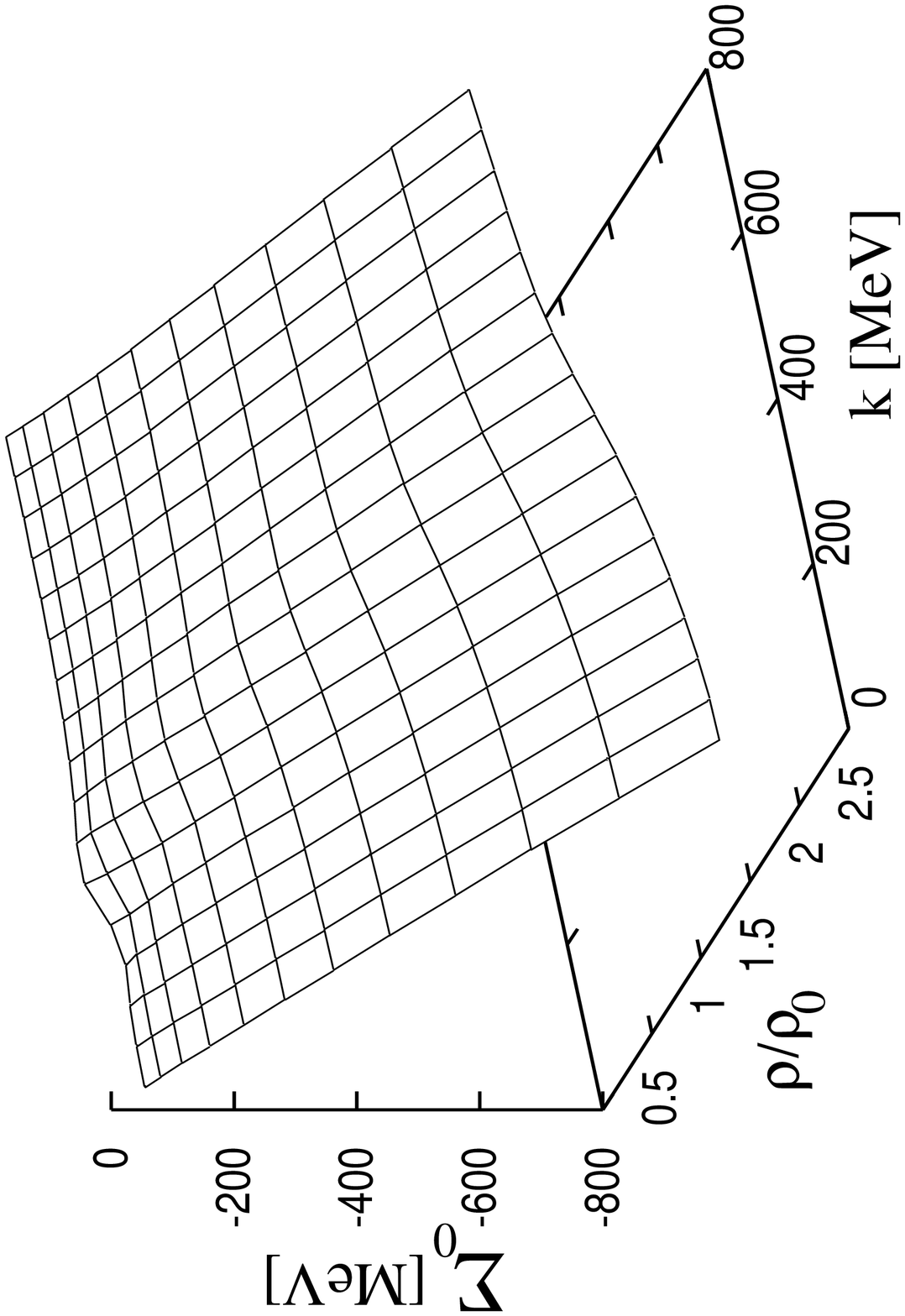}
\caption{\label{fig10} Dependence of the nucleon 
self-energy component $\Sigma_s$ (upper panel) and $\Sigma_0$ 
(lower panel) on the nuclear matter density $\rho$ and 
the nucleon momentum $k$. As bare interaction the Bonn A potential is used 
while for the T-matrix the subtraction scheme with the $ps$ representation,
Eq. (\ref{tmatps}), is applied.} 
\end{figure}

In Fig. \ref{fig10} we show the complete dependence of the 
scalar $\Sigma_S$ self-energy and the vector component  
$\Sigma_0$ on momentum and density. 
As bare interaction the Bonn A potential is used again while for the 
T-matrix the subtraction scheme with the $ps$ representation  
of the ladder kernel, Eq. (\ref{tmatps}), is applied. 
As can be seen from Fig. \ref{fig10} the momentum dependence 
starts to become pronounced at densities around
$\rho=2.0 \rho_0$, where $\rho_0=0.166 fm^{-3}$ is the empirical 
saturation density of nuclear matter.
At fixed density the momentum dependence is still most pronounced around the
corresponding Fermi momentum. On the other hand, keeping the nucleon
momentum fixed and varying the Fermi momentum, we see that the medium dependence,
namely the variation of the self-energy with the nuclear matter density, is
strongest for low energetic nucleons. This clearly demonstrates the influence of
the Pauli-blocking effect which vanishes with increasing relative momentum of the
nucleon interacting with the particles inside the Fermi sea.
\subsubsection{Density dependence}
\begin{figure}[h]
\vspace{75mm}
\includegraphics{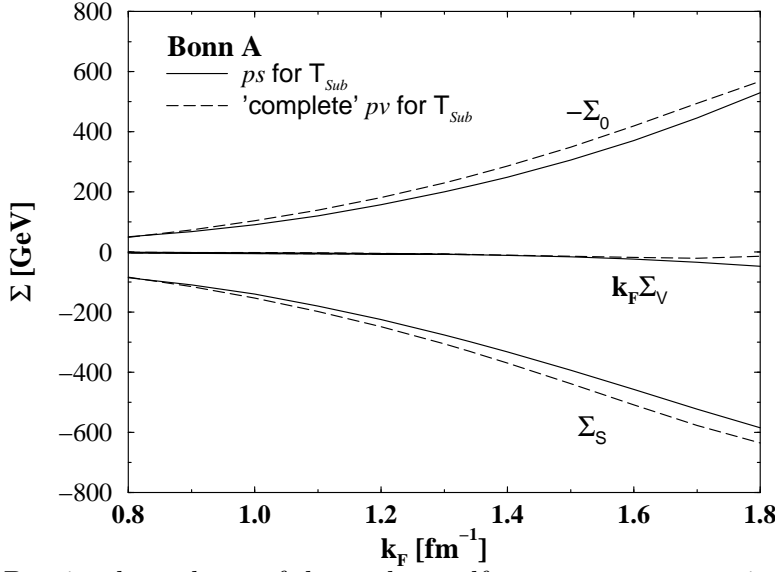}
\caption{\label{fig11} Density dependence of the nucleon self-energy
components in nuclear matter using Bonn A as bare nucleon-nucleon interaction.
For the T-matrix the subtraction scheme with the 
$ps$ representation (Eq. (\ref{tmatps}), solid) and the 
$pv$ representation (Eq. (\ref{tmatpv}), dashed) are applied.}
\end{figure}

The detailed density dependence of the nucleon self-energy components in 
nuclear matter is presented in Fig. \ref{fig11}. 
The momentum k of the nucleon is thereby fixed at the Fermi momentum $k_F$. 
The density dependence of the nucleon self-energy is quite similar for 
both representations (\ref{tmatps}) and (\ref{tmatpv}) of the T-matrix. 
An important difference is, however, the deviation in the vector component $\Sigma_V$. 
As already seen from Fig. \ref{fig9}, the 'complete' $pv$ representation of the
ladder kernel leads to an unphysical behavior of the spatial
$\Sigma_V$ component. 

\begin{figure}[h]
\vspace{75mm}
\includegraphics{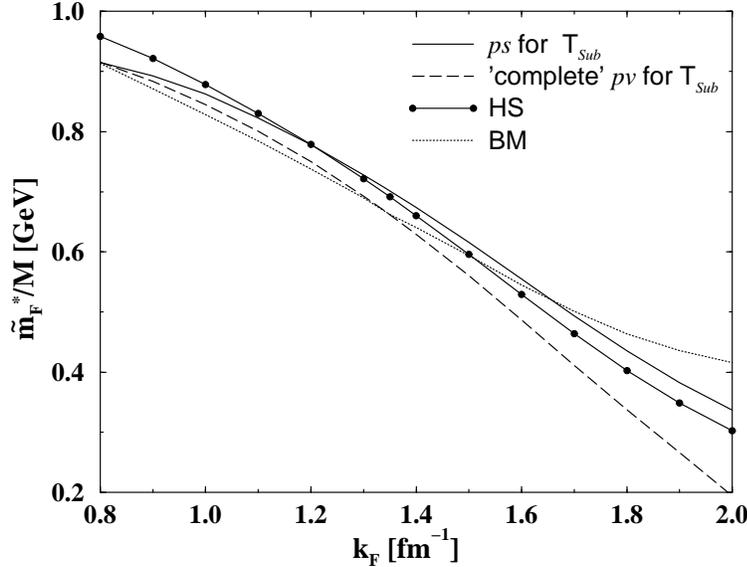}
\caption{\label{fig12} Density dependence of the reduced effective mass 
$\tilde{m}_F^*$ of the nucleon in nuclear matter, 
using Bonn A as bare nucleon-nucleon 
interaction. For the T-matrix the subtraction scheme with 
the $ps$ representation (Eq. (\ref{tmatps}), solid) and the 
$pv$ representation (Eq. (\ref{tmatpv}), dashed) are applied. 
In addition the result of Horowitz and Serot (HS)  
(Ref. \protect\cite{horowitz87}, solid with circles) and 
of Brockmann and Machleidt (BM) (Ref. \protect\cite{brockmann90}, dotted) 
are shown.}
\end{figure}

This has a strong influence on the reduced 
effective mass $\tilde{m}_F^*$ of the nucleon, Eq. (\ref{redquantity}). 
The reduced effective mass generally drops with increasing density as 
can be seen from Fig. \ref{fig12} where the reduced effective mass is shown  
as a function of the Fermi momentum. 
At saturation density, for Bonn A at $k_F=1.39 fm^{-1}$, the reduced
effective mass has a value of $\tilde{m}_F^* \sim 0.65 M $ 
for both T-matrix representations. This agrees with the findings of
other groups \cite{horowitz87,brockmann90} 
and also with the empirical value determined from the spin-orbit interaction 
in finite nuclei \cite{ring96}.
At larger densities, however, the two representations for the T-matrix lead 
to rather different results.
At a Fermi momentum of $k_F=2.0 fm^{-1}$, which is three times nuclear 
matter density, the results differ by almost a factor of two. 
The solid curve with circles shown in Fig. \ref{fig12} is the result of the 
calculation by Horowitz and Serot \cite{horowitz87}, who used as 
bare nucleon-nucleon interaction the $\sigma$-$\omega$ model potential. 
Since they did not consider the pion in their calculation they used the
$ps$ representation for the full in-medium T-matrix. 
Fitting the saturation properties of nuclear matter, their result for the 
effective mass is almost identical to our calculation with 
the $ps$ representation (\ref{tmatps}) of the subtracted T-matrix. 
The calculation of Brockmann and Machleidt \cite{brockmann90} 
is shown as dotted curve in Fig. \ref{fig12}.  
These authors used instead of the projection technique 
a fit procedure to the single-particle potential. They 
assumed a momentum independent form of the nucleon self-energy and 
thus they could obtain only approximately the effective mass of the nucleon. 
For lower densities their result is, however, quite
similar to our findings. Only at higher densities important
differences occur. In their calculation 
the effective mass seems to saturate much earlier 
while in our calculation with the $ps$ representation of the 
subtracted T-matrix the effective mass drops to a smaller value 
at larger densities.

The above analysis indicates again that the $ps$ representation of the 
remaining ladder kernel of the T-matrix is preferable compared to the 
'complete' $pv$ representation.
Although the two alternatives proposed here yield similar results at densities 
below and up to saturation, the different representations 
become decisive with increasing density. 
The 'complete' $pv$ representation of the ladder kernel of the T-matrix 
leads thereby to a partially unphysical high density behavior, i.e. a
too strongly dropping mass, whereas the $ps$
representation is still reasonable.
For completeness the various relevant quantities as a function of the
nuclear matter density are presented in Tab. \ref{table1} 
using the subtraction scheme with 
the $ps$ representation (\ref{tmatps}) for the subtracted T-matrix. As
bare nucleon-nucleon interaction the Bonn A potential is used. 
\begin{table}
\begin{center}
\caption{
The Fermi-momentum $\kf$, the nuclear matter density $\rho$, 
the binding energy per particle $E/A$, the
reduced effective mass $\tilde{m}^*_F$ and the components of the
nucleon self-energy (at $|{\bf k}|=\kf$) 
for nuclear matter applying the $ps$ representation 
(\ref{tmatps}) for the subtracted T-matrix. 
As bare nucleon-nucleon interaction the Bonn A potential is used.}
\begin{tabular}{ccccccc}
\\
\hline\hline 
$\kf$ & $\rho$ & $E/A$  &  $\tilde{m}^*_F$ & $\Sigma_s$ & $-\Sigma_0$ & $\Sigma_v$ \\
$[{\rm fm}^{-1}]$ & $[{\rm fm}^{-3}]$ &  [MeV]  &  [MeV]   &  [MeV]    & [MeV]    &   \\
\hline
0.7 & 0.023 & -7.74  &  868.4  &  -72.0   &    42.1    &   -0.0017  \\   
0.8 & 0.035 & -8.77  &  858.2  &  -85.6   &    51.1    &   -0.0058   \\   
0.9 & 0.049 & -9.98  &  837.1  &  -108.8  &    68.0    &   -0.0084   \\   
1.0 & 0.068 & -11.62 &  809.1  &  -139.5  &    90.6    &   -0.0121   \\   
1.1 & 0.090 & -13.45 &  771.6  &  -179.2  &    120.9   &   -0.0155   \\  
1.2 & 0.117 & -14.80 &  729.9  &  -224.0  &    157.3   &   -0.0207   \\  
1.3 & 0.148 & -15.74 &  682.8  &  -275.2  &    200.3   &   -0.0281   \\  
1.35& 0.166 & -16.03 &  657.8  &  -303.0  &    224.2   &   -0.0333   \\  
1.4 & 0.185 & -16.15 &  632.1  &  -331.8  &    249.5   &   -0.0395   \\  
1.5 & 0.228 & -15.28 &  577.8  &  -392.6  &    305.8   &   -0.0546   \\  
1.6 & 0.277 & -12.37 &  520.4  &  -457.0  &    371.1   &   -0.0741   \\  
1.7 & 0.332 & -6.19  &  462.9  & -522.2   &    445.8   &   -0.0999    \\ 
1.8 & 0.394 &  4.78  &  408.4  & -584.6   &    530.2   &   -0.1325    \\ 
1.9 & 0.463 & 22.22  &  358.9  & -641.4   &    626.4   &   -0.1712    \\ 
2.0 & 0.540 & 48.19  &  315.9  & -690.8   &    735.6   &   -0.2146    \\ 
\hline
\end{tabular}
\label{table1}
\end{center}
\end{table}
\subsection{The equation-of-state of nuclear matter}
\begin{figure}[h]
\vspace{75mm}
\includegraphics{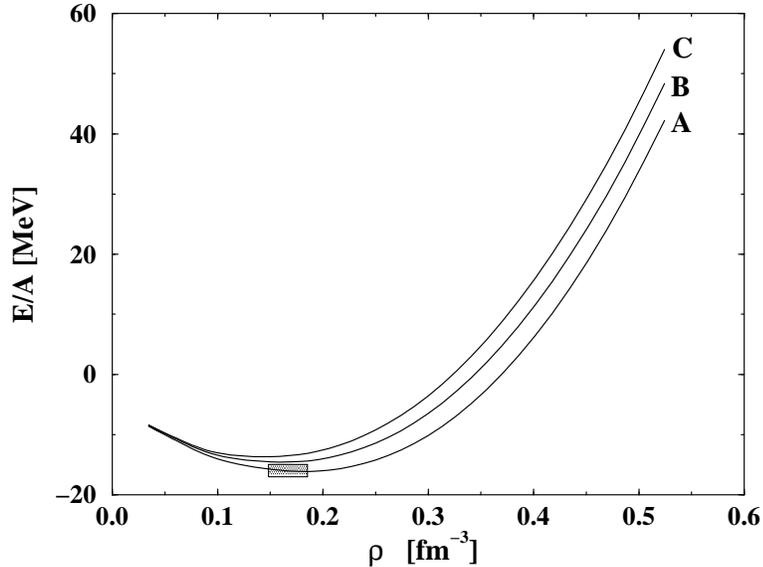}
\caption{\label{fig13} Binding energy per particle as a 
function of nuclear matter density. 
As bare nucleon-nucleon interaction the potentials 
Bonn A, B and C are used. For the T-matrix the subtraction scheme with the 
$ps$ representation (\ref{tmatps}) is applied. 
The shaded box denotes the empirical saturation region of nuclear matter.}
\end{figure}

In the relativistic Brueckner theory the energy per particle is defined 
as the kinetic plus half the potential energy 
\beq
E/A  =  {1\over \rho}\sum_{{\bf k},\lambda}
<\bar{u}_\lambda({\bf k})| 
\bfgamma \cdot {\bf k} + M + {1\over 2}\Sigma (k)
| u_\lambda({\bf k})>
{{\tilde{m}^*(k)}\over {{\tilde{E}}^*(k)}} - M
\quad .
\label{eos}
\eeq
In Fig. \ref{fig13} we show the binding energy per particle $E/A$ as a 
function of the density, calculated with Bonn A, B and C.
For the T-matrix the subtraction scheme with the $ps$ representation 
(\ref{tmatps}) is applied.
With Bonn A one can reproduce the empirical saturation point
of nuclear matter, shown as shaded region in the figure.
The other Bonn potentials give less binding energy although the 
saturation density is always close to the empirically known value.

\begin{figure}[h]
\vspace{75mm}
\includegraphics{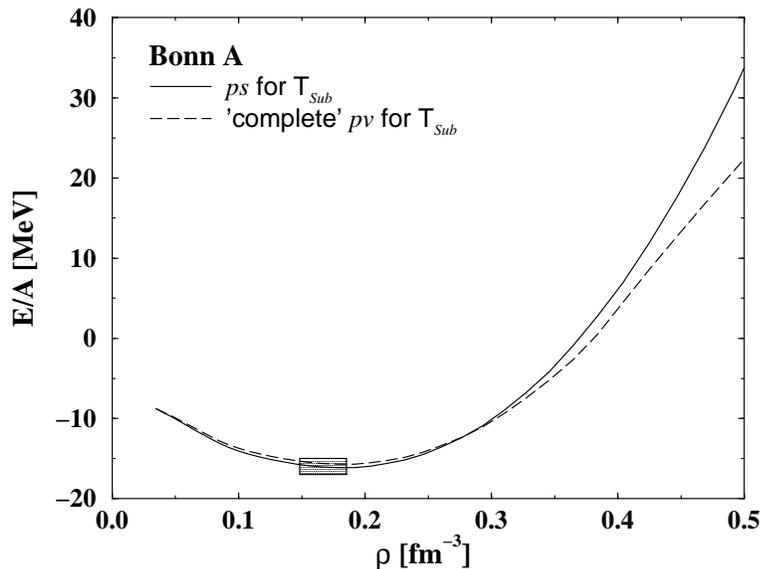}
\caption{\label{fig14} Binding energy per particle as a 
function of nuclear matter density. As bare nucleon-nucleon interaction 
the Bonn A potential is used. For the T-matrix the subtraction scheme with 
the $ps$ representation (Eq. (\ref{tmatps}), solid) and the 
$pv$ representation (Eq. (\ref{tmatpv}), dashed) are applied.}
\end{figure}

The result for the binding energy per particle using the two 
representations (\ref{tmatps}) and (\ref{tmatpv}) for the T-matrix 
are very similar as can be seen in Fig. \ref{fig14}.
At saturation density the binding energy is only 0.5 MeV smaller using the 
pseudo-vector representation of the subtracted T-matrix. 
Thus, the energy per particle is not very sensitive on the explicit representation of the 
subtracted T-matrix. As already noticed, on the level of 
the self-energies, Fig. \ref{fig5}, \ref{fig9} and
\ref{fig12}, differences between the two methods occur at higher densities.
In particular the equation-of-state in the $ps$ representation for the subtracted
T-matrix appears to be more stiff at higher densities than with the corresponding
$pv$ representation. 

\begin{figure}[h]
\vspace{75mm}
\includegraphics{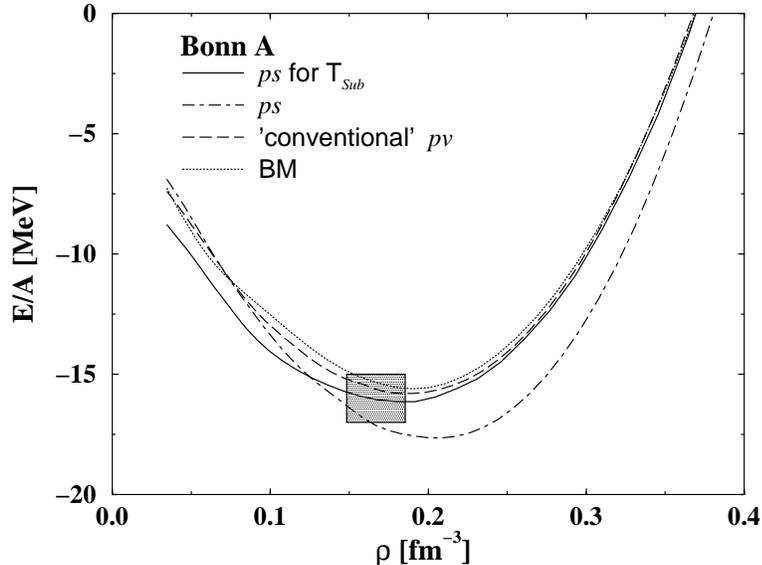}
\caption{\label{fig15} Binding energy per particle as a 
function of nuclear matter density. 
As bare nucleon-nucleon interaction the Bonn A potential 
is used. For the T-matrix the subtraction scheme with the $ps$ representation 
(Eq. (\ref{tmatps}), solid), the full $ps$ representation 
(Eq. (\ref{tmatrep1}), dash-dotted) and the 'conventional' $pv$ representation
(see text after Eq. (\ref{tmatidentity}), dashed) are applied. 
In addition the result of Brockmann and Machleidt (BM) 
(Ref. \protect\cite{brockmann90}, dotted) is shown.}
\end{figure}

To understand this behavior in more detail and to compare also  
with other calculations we show in Fig. \ref{fig15} the binding energy
per particle as a function of nuclear matter density for different scenarios. 
First of all, applying the projection method of Horowitz and Serot, we
can reproduce almost completely the result of Brockmann and Machleidt (BM)
, Ref. \cite{brockmann90}, 
when we use the 'conventional' $pv$ representation of the T-matrix. 
(see text after Eq. (\ref{tmatidentity})). The two approaches are 
in detail compared in Tab. 3. Except of the incompressibility, 
which is generally smaller in our calculation, the bulk properties are 
indeed very similar. This result is somewhat surprising since the 
approach of Ref. \cite{brockmann90} is completely different from 
the present one. As already mentioned, in Ref. \cite{brockmann90} no 
projection scheme to the T-matrix has been applied but constant, i.e. 
momentum independent, self-energy components have been determined by 
a fit to the single particle potential. On the other hand, the 
'conventional' $pv$ representation of the T-matrix leads to a 
pronounced momentum dependence of the self-energy components, see 
Fig. 4. That these two calculations give nevertheless similar results 
for the nuclear matter bulk properties is due to the fact that they 
lead to more or less identical values for the fields and the 
effective mass at the Fermi momentum. This agreement appears, however, 
to be somewhat fortuitous.

However, as discussed in the previous sections the 
'conventional' $pv$ representation 
does not correctly reproduce the contribution of the 
single-pion exchange potential to the nucleon self-energy. 
On the level of the binding energy one can estimate the contribution of the 
single-pion exchange potential comparing the result of 
$ps$ representation for the subtracted T-matrix and the 
pure $ps$ representation of the T-matrix. 
In the latter approach the nucleons are less bound at small 
densities but the situation changes around saturation. 
A correct pseudo-vector representation
of the pion, as used in the subtraction scheme,  
suppresses this effect. Thus at smaller densities we obtain a larger
binding, while around saturation density the binding energy is smaller.
Compared to the 'conventional' $pv$ representation 
or the result of Brockmann and Machleidt, the 
$ps$ representation of the subtracted T-matrix plus a correct treatment of
the bare interaction leads altogether to more binding at smaller and medium densities. 
The correct treatment of the single-pion exchange potential is therefore  
essential in the low density regime of the equation-of-state. 
Except of the pure $ps$ treatment, which is certainly not correct for a realistic
potential like the Bonn potential \cite{nuppenau89,terhaar87,fuchs98}, 
the various calculations coincide at high
densities. This reflects a decreasing relative 
importance of the pion-exchange at high
densities. 

\begin{figure}[h]
\vspace{75mm}
\includegraphics{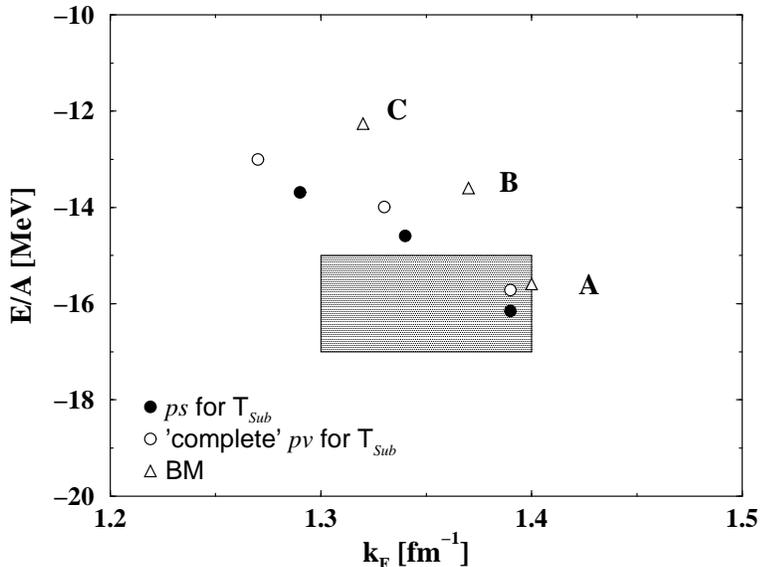}
\caption{\label{fig16} Saturation points of nuclear matter. As bare 
nucleon-nucleon interaction the Bonn potentials A,B and C are used.
For the T-matrix the subtraction scheme with the 
$ps$ representation (Eq. (\ref{tmatps}), filled circles) 
and the $pv$ representation (Eq. (\ref{tmatpv}), open circles) 
are applied. As open triangles the results of the calculation of 
Brockmann and Machleidt (BM), Ref. \protect\cite{brockmann90}, are shown.}
\end{figure}

The present results are summarized in Fig. \ref{fig16} 
where the corresponding saturation points for the three 
different versions of the Bonn potential are shown. 
We compare the results with the two representation of the
subtracted T-matrix with the results of the calculation of 
Brockmann and Machleidt (BM), Ref. \cite{brockmann90}.
With the improved representation schemes 
(\ref{tmatps}) and (\ref{tmatpv}) for the T-matrix one obtains new 
'Coester-lines' which are left of the original one, 
i.e. shifted towards the empirical region. The refined treatment of the 
T-matrix representation leads to an enhancement
of the binding energy connected with a reduced saturation density. As
in the previous calculations, Bonn A is still the only one which 
meets the empirical region. However, due to an increased binding 
energy Bonn B is now much closer to empirical region. This observation 
is consistent with the present treatment. The different types of Bonn 
potentials essentially vary in the strength of the nuclear tensor 
force determined by the $\pi NN$ form factor. Bonn A which has the 
smallest tensor force yields the smallest D-state probability of 
the deuteron and only a pure description of the 
$^{3}D_{1}$ phase shift \cite{brockmann90,machleidt86}. 
Thus it appears that a refined treatment of the 
pion exchange leads to improved nuclear matter results for the 
more realistic Bonn B potential. Bonn C, however, is still 
far off the empirical region.

Furthermore, it can be seen from Fig. 16 and 
Tab. 2 that the final nuclear matter bulk properties depend 
only moderate on the representation of the 
subtracted T-matrix. In Ref. \cite{fuchs98} we tried already to 
determine the range of inherent uncertainty in the relativistic 
Brueckner approach which is due to the ambiguities concerning the 
representation of the T-matrix discussed in Section 3. 
By the separate treatment of the Born contribution to the T-matrix 
we end up now with a much narrower uncertainty band which is 
given by the $ps$ or complete $pv$ representation of the ladder 
kernel, i.e. the subtracted T-matrix. Over the different types 
of Bonn interactions the two methods lead to a variation of 
0.5 MeV in the binding energy, 0.1--0.2 $fm^{-1}$ in the Fermi 
momentum, and to about 30 MeV concerning the value of the 
effective mass at saturation density. The values for 
incompressibility are also close in the two approaches, i.e. they 
differ by less than 10 MeV.  
However, compared to the conventional $pv$ representation 
and especially to Ref. \cite{brockmann90}, see Tab. \ref{table3}, 
the kompression moduli are significantly reduced for all three types of 
interactions when we use the new approach to the T-matrix representation, see
Tab. \ref{table2}. Within the $ps$ representation of the subtracted T-matrix, 
Bonn B and C now yield very small kompression moduli around
$K=150 MeV $ and $K=115 MeV$, respectively. 
For Bonn A a value of $K=230 MeV$ is obtained. 
This value agrees with the empirical value of the 
kompression modulus of $K=210 \pm 30 MeV$ \cite{blaizot80}.
Here Brockmann and Machleidt found much larger values for all three 
Bonn potentials. 

\begin{table}
\begin{center}
\caption{
The Fermi-momentum $\kf$, the binding energy per particle $E/A$, the
reduced effective mass $\tilde{m}^*_F$ and the kompression modulus
K at saturation for nuclear matter using as bare nucleon-nucleon interaction the 
Bonn potentials A, B and C. 
For the T-matrix the subtraction scheme with 
the $ps$ representation (\ref{tmatps}) and the 
$pv$ representation (\ref{tmatpv}) are applied.}
\begin{tabular}{cccccccccc}
\\
\hline\hline 
& \multicolumn{4}{c}{{\it ps} for ${\rm T}_{Sub}$} &  
& \multicolumn{4}{c}{'complete'{\it pv} for ${\rm T}_{Sub}$} \\
\cline{2-5} \cline{7-10} 
      & $\kf$       &   $E/A$    &  $\tilde{m}^*_F$ &   K  & 
      & $\kf$       &   $E/A$    &  $\tilde{m}^*_F$ &   K  \\
      & [fm$^{-1}$] & [MeV]  & [MeV]   & [MeV] &
      & [fm$^{-1}$] & [MeV]  & [MeV]   & [MeV] \\
\hline
 A         & 1.39        & -16.15  &  637.0    & 230 &
           & 1.39        & -15.72  &  596.0    & 220 \\
 B         & 1.34        & -14.59  &  667.0    & 150 & 
           & 1.33        & -13.99  &  634.0    & 140 \\             
 C         & 1.29        & -13.69  &  694.0    & 110 &
           & 1.27        & -13.00  &  669.0    & 100 \\
\hline
\end{tabular}
\label{table2}
\end{center}
\end{table}
\begin{table}
\begin{center}
\caption{
The Fermi-momentum $\kf$, the binding energy per particle $E/A$, the
reduced effective mass $\tilde{m}^*_F$ and the kompression modulus
K at saturation for nuclear matter using as bare nucleon-nucleon interaction the 
Bonn potentials A, B and C. For the T-matrix the 
'conventional' $pv$ representation, see text after Eq. (\ref{tmatidentity}), is
applied. In addition the results of Brockmann and Machleidt (BM),
Ref. \protect\cite{brockmann90}, are presented.}
\begin{tabular}{cccccccccc}
\\
\hline\hline 
& \multicolumn{4}{c} {'conventional' $pv$} &  
& \multicolumn{4}{c} {BM}  \\
\cline{2-5} \cline{7-10} 
      & $\kf$       &   $E/A$    &  $\tilde{m}^*_F$ &   K  & 
      & $\kf$       &   $E/A$    &  $\tilde{m}^*_F$ &   K  \\
      & [fm$^{-1}$] & [MeV]  & [MeV]   & [MeV] &
      & [fm$^{-1}$] & [MeV]  & [MeV]   & [MeV] \\
\hline
 A         & 1.41        & -15.81  &  538.0    & 275 &
           & 1.40        & -15.59  &  564.0    & 290 \\
 B         & 1.35        & -13.70  &  565.0    & 195 & 
           & 1.37        & -13.60  &  573.0    & 249 \\             
 C         & 1.30        & -12.31  &  585.0    & 155 &
           & 1.32        & -12.26  &  590.0    & 185 \\
\hline
\end{tabular}
\label{table3}
\end{center}
\end{table}
\clearpage
\section{Summary}
We have investigated the nuclear matter properties in the 
relativistic Brueckner approach. The required representation of 
the T-matrix by Lorentz invariant amplitudes suffers thereby 
from on-shell ambiguities concerning the pseudo-scalar or 
pseudo-vector nature of the interaction. We minimized this ambiguity 
by separating the leading order, i.e. the single-meson exchange,  
from the full T-matrix. Actually we represented the contributions 
stemming from the single-meson exchange by taking the pseudo-scalar
and the pseudo-vector nature of the interaction into account. Up to now, 
this approach is the only one which reproduces the correct results 
for the T-matrix on the Hartree-Fock level. 
The remaining higher order correlations, i.e. the ladder kernel, are then 
represented either completely as pseudo-scalar or as pseudo-vector. 

This method takes at best the pion contribution to the nucleon 
self-energy into account which, on the other hand, essentially determines 
the momentum dependence of the self-energy. Treating the one-pion 
exchange with a pseudo-vector coupling the momentum dependence 
is strongly suppressed as it is desired from meson phenomenology. This 
also favors the 'reference spectrum approximation' which was
a first guess to the momentum dependence of the effective mass of the nucleon.
Now the results of the calculation are much more consistent with the 
initial assumption of a weakly momentum dependent effective mass. 

Furthermore, we have investigated the density dependence of the nucleon self-energy.
The results for both scenarios for the T-matrix representation are quite similar 
up to saturation density. This holds also for the binding energy. 
Both representations lead to similar 
saturation properties of nuclear matter which indicates that the 
remaining ambiguity in the representation of the ladder kernel, i.e. 
the higher order correlations, is not too severe at moderate densities. 
However, the two schemes start to differ significantly at higher 
densities. There, 
the complete $pv$ representation of the ladder kernel, i.e.
the subtracted T-matrix, leads to an unphysical behavior, 
whereas the corresponding
$ps$ representation still leads to reasonable results. 
Thus it appears that the higher density correlations 
are best represented adopting the
pseudo-scalar representation.

As a major result of our investigation we obtain new 'Coester lines' 
for the various Bonn potentials. Compared to previous treatments 
these are shifted towards
the empirically know saturation point. Bonn A is still the only potential
which really meets the empirical region of saturation, but, 
with improved saturation properties compared to 
previous treatments. The refined treatment of the pion exchange 
leads on the other hand also to improved results for 
the -- from the view of the phase shift analysis -- more realistic 
Bonn B potential. Furthermore, we found that the 
equation-of-state is strongly softened compared to previous calculations.
Actually with Bonn A we obtain a kompression modulus of $K\sim 230 MeV$ 
which is in good agreement with the empirical value.

To summarize our results, we obtained new results for the nuclear matter
properties within the projection technique employing an new method for the 
T-matrix representation. The final results are at lower
densities almost insensitive on the explicit choice made for the 
representation. However, at higher densities, certain differences occur when 
using different representation schemes. We want to stress that the ambiguity in the
projection technique is still not fully resolved yet. We plan to look on off-shell
T-matrix elements in the future since off-shell matrix elements of the 
pseudo-scalar and pseudo-vector covariants differ significantly. 
We hope that this might bring more insight on what is the correct on-shell 
representation of the T-matrix. 
\begin{ack}
The authors would like to thank H. M\"uther and E. Schiller 
for valuable discussions. 
\end{ack}


\begin{thebibliography}{99} 
\bibitem{horowitz87}
C.J. Horowitz and B.D. Serot,
Nucl. Phys. A464 (1987) 613.

\bibitem{terhaar87}
B. ter Haar and R. Malfliet,
Phys. Rep. 149 (1987) 207.

\bibitem{brockmann90}
R. Brockmann, R. Machleidt,
Phys. Rev. C42 (1990) 1965.

\bibitem{sehn97}
L. Sehn, C. Fuchs and A. Faessler,
Phys. Rev. C56 (1997) 216.

\bibitem{fuchs98}
C. Fuchs, T. Waindzoch, A. Faessler and D.S. Kosov,
Phys. Rev. C58 (1998) 2022.

\bibitem{dejong98}
F. de Jong, H. Lenske,
Phys. Rev. C58 (1998) 890.

\bibitem{coester70}
F. Coester, S. Cohen, B.D. Day and C.M. Vincent, 
Phys. Rev. C1 (1970) 769.

\bibitem{erkelenz74}
K. Erkelenz,
Phys. Rep. C13 (1974) 191.

\bibitem{holinde81}
K. Holinde,
Phys. Rep. C68 (1981) 121.

\bibitem{nuppenau89}
C. Nuppenau, Y.J. Lee and A.D. MacKellar,
Nucl. Phys. A504 (1989) 839.

\bibitem{lee97}
C.-H. Lee, T.S. Kuo, G.Q. Li and G.E. Brown,
Phys. Lett. B412 (1997) 235.

\bibitem{machleidt86}
R. Machleidt,
Advances in Nuclear Physics, 19, 189, eds. J.W. Negele, E. Vogt 
(Plenum, N.Y., 1986).

\bibitem{tjon85b}
J.A. Tjon and S.J. Wallace,
Phys. Rev. C32 (1985) 1667.

\bibitem{weinberg68}
S. Weinberg,
Phys. Rev. Lett. 116 (1968) 1568.

\bibitem{terhaar87b}
B. ter Haar and R. Malfliet,
Phys. Rev. 36 (1987) 1611.

\bibitem{fritz94}
R. Fritz and H. M\"uther,
Phys. Rev. 49 (1994) 633.

\bibitem{brown76}
G.E. Brown and A.D. Jackson,
The Nucleon-Nucleon Interaction (North Holland, Amsterdam, 1976).

\bibitem{thompson70}
R.H. Thompson,
Phys. Rev. D1 (1970) 110.

\bibitem{serot86}
B.D. Serot and J.D. Walecka,
Advances in Nuclear Physics, 16, 1, eds. J.W. Negele, E. Vogt (Plenum, N.Y., 1986).

\bibitem{bethe63}
H.A. Bethe, B.H. Brandow and A.G. Petschek,
Phys. Rev. 129 (1963) 225.

\bibitem{anastasio83}
M.R. Anastasio, L.S. Celenza, W.S. Pong and C.M. Shakin,
Phys. Rep. 100 (1983) 327.

\bibitem{poschenrieder88}
P. Poschenrieder, M.K. Weigel,
Phys. Rev. C38 (1988) 471.

\bibitem{sehn98}
L. Sehn, A. Faessler and C. Fuchs,
J. Phys. G.: Nucl. Part. Phys. 24 (1998) 135.

\bibitem{haftel70}
M.I. Haftel and F. Tabakin,
Nucl. Phys. A158 (1970) 1.

\bibitem{rose57}
M. Rose,
Elementary Theory of Angular Momentum (Wiley, N.Y., 1957).

\bibitem{tjon85a}
J.A. Tjon and S.J. Wallace,
Phys. Rev. C32 (1985) 267.

\bibitem{kondratyuk98}
S. Kondratyuk and O. Scholten,
nucl-th/9807077.

\bibitem{boersma94}
H.F. Boersma and R. Malfliet,
Phys. Rev. C49 (1994) 233.

\bibitem{walecka74}
J.D. Walecka,
Ann. Phys. (N.Y.) 83 (1974) 497.

\bibitem{ring96}
P. Ring, 
Prof. Part. Nucl. Phys. 37 (1996) 137.

\bibitem{blaizot80}
J.P. Blaizot,
Phys. Rep. 65 (1980) 171.

\end{thebibliography}
\end{document}